%% file: ScientificData.tex
\documentclass[fleqn,10pt]{wlscirep}
\usepackage[utf8]{inputenc}
\usepackage[T1]{fontenc}

\usepackage{lineno}
\nolinenumbers 

\usepackage{tikz}
\usepackage{mathtools}
\usepackage{tabularray}
\usepackage{tabularx,calc,array}
\usepackage{amssymb}
\usepackage{float}
\usepackage{amsfonts}
\usepackage{geometry}
\usepackage{graphicx} 
\usepackage[skip=0.333\baselineskip]{caption}
\usepackage{subcaption}

\usepackage [english]{babel}
\usepackage [autostyle, english = american]{csquotes}
\MakeOuterQuote{"}

\usepackage{booktabs}
\usepackage{multirow}

\title{A Dataset of Uniswap daily transaction indices by network}

\author[a]{Nir Chemaya}
\author[b,*]{Lin  William  Cong}
\author[a]{Emma Jorgensen}
\author[a,*]{Dingyue Liu}
\author[c,*]{Luyao Zhang}
\affil[a]{University of California, Santa Barbara, Santa Barbara, 93106, USA}
\affil[b]{Cornell SC Johnson College of Business, IC3, \& NBER, Ithaca, NY, 14853, USA}
\affil[c]{Duke Kunshan University, 8 Duke Avenue, Kunshan, Jiangsu, 215316, China}

\affil[*]{Contact Lin William Cong (will.cong@cornell.edu) at Cornell University, Dingyue Liu (dingyueliu@ucsb.edu) at the University of California, Santa Barbara, Santa Barbara, and Luyao Zhang (lz183@duke.edu) at Data Science Research Center and Social Science Division, Duke Kunshan University. \newline The authors are by the alphabetical order of their last names.}

\begin{abstract}

\input{0.Abstract}

\end{abstract}

\begin{document}

\flushbottom
\maketitle

\input{1.backgroundsummary}

\input{2.methods}

\input{3.datarecord_validation}
\input{4.UsageNotes}
\input{5.codeavailability}
\input{6.acknowledgements}
\input{7.figures_tables}

\thispagestyle{empty}



\clearpage

\bibliography{newrefs}

\clearpage

\input{8.appendix}

\end{document}

%% file: 0.Abstract.tex
Decentralized Finance (DeFi) is reshaping traditional finance by enabling direct transactions without intermediaries, creating a rich source of open financial data. Layer 2 (L2) solutions are emerging to enhance the scalability and efficiency of the DeFi ecosystem, surpassing Layer 1 (L1) systems. However, the impact of L2 solutions is still underexplored, mainly due to the lack of comprehensive transaction data indices for economic analysis. This study bridges that gap by analyzing over 50 million transactions from Uniswap, a major decentralized exchange, across both L1 and L2 networks. We created a set of daily indices from blockchain data on Ethereum, Optimism, Arbitrum, and Polygon, offering insights into DeFi adoption, scalability, decentralization, and wealth distribution. Additionally, we developed an open-source Python framework for calculating decentralization indices, making this dataset highly useful for advanced machine learning research. Our work provides valuable resources for data scientists and contributes to the growth of the intelligent Web3 ecosystem.

%% file: 1.backgroundsummary.tex
\section*{Background and Summary}
The advent of Decentralized Finance (DeFi) has wrought a paradigm shift within the financial sector by introducing a novel approach to financial services and intermediation.~\cite{harvey2021defi} DeFi offers a spectrum of on-chain services, including trading, borrowing, lending, and investing, all of which circumvent traditional intermediaries.~\cite{zhang2023blockchain,malinova2023learning,ao2022decentralized,petryk2022impact,lehar2021decentralized, appel2023decentralized,lehar2022systemic} At the heart of this transformative ecosystem lie smart contracts, programmable software applications that facilitate the automation of financial transactions\cite{Auer2023}. These smart contracts are deployed on blockchain networks, with Ethereum emerging as the dominant platform for DeFi activities. Nonetheless, alternative platforms like Polygon also provide similar functionalities, contributing to the diversification of the DeFi landscape. The allure of DeFi has burgeoned in recent years, with the Total Value Locked (TVL) in DeFi protocols soaring to an estimated $163$ billion USD as of January 1, 2022.\cite{defillama} Ethereum, with its robust infrastructure, commands a substantial share, accounting for 56\% of the TVL. Other blockchain networks contend for the remaining market share. Amidst this fervor, the pursuit of decentralization, security, and scalability within the blockchain realm has come to the fore, where decentralization refers to reducing reliance on central authorities, security ensures the protection of the network and data integrity, and scalability reflects the system’s capacity to handle increasing transaction volumes efficiently.

However, the blockchain trilemma\cite{bisS22, hafid2020scaling} underscores the intrinsic trade-offs among these attributes. Achieving all three concurrently remains an elusive goal. Ethereum, for instance, prioritizes security and decentralization but grapples with scalability challenges. Its transaction processing capacity typically ranges from 15 to 30 transactions per second, paling in comparison to traditional payment systems like Visa, which handle up to 60,000 transactions per second. Furthermore, Ethereum's scalability constraints often result in elevated gas fees for transaction validation, especially during network congestion. These limitations also have implications for inclusion, equality, and democratization.\cite{makarov2022cryptocurrencies,cong2023inclusion} Against this backdrop, Layer 2 (L2) blockchains have surfaced, promising expedited transaction speeds and reduced fees compared to Layer 1 (L1) counterparts like Ethereum, on top of which they are built to enhance scalability and efficiency.\cite{cong2023scaling} Emerging L2 blockchains, exemplified by Polygon, boast the capacity to process thousands of transactions per second, offering a compelling alternative to users. Network fees, particularly during congested periods, significantly influence users' decisions regarding blockchain transactions.\cite{easley2019mining} The availability of L2 blockchains as a more cost-effective and efficient network alternative may drive users to migrate their activities to L2 environments.

Nonetheless, the precise ramifications of L2 solutions remain shrouded in uncertainty, primarily due to the paucity of transaction data indices that furnish meaningful economic insights for empirical analysis. This research endeavors to bridge this knowledge gap by embarking on an exhaustive analysis of raw transaction data harvested from Uniswap, a pivotal decentralized exchange (DEX) and linchpin of the DeFi ecosystem, which many researchers have already explored both theoretically and empirically.~\cite{barbon2021quality,angeris2019analysis,park2023conceptual,lehar2021decentralized,xia2021trade,lo2022uniswap,chemaya2023default,chemaya2023suitability,chemaya2023estimating}
In Figure ~\ref{fig:transaction_mindmap}, we present an illustrative depiction of how transaction data is systematically recorded within the Uniswap ecosystem. To elucidate this process, consider a hypothetical user named Alice, who engages in a trade involving a specific quantity of token Xi for token Y. This transaction takes place within a liquidity pool containing both token X and token Y. The intricacies of this transaction are securely recorded on the blockchain, allowing us to extract a wealth of transaction-related information.

Within this dataset, several key attributes stand out. The timestamp denotes the precise moment when the trade was executed, providing a temporal reference point. The wallet address offers insight into the pseudonymous identity of the user responsible for executing the trade. The pool ID furnishes vital details about the pairs of cryptographic tokens available for trading within the pool, encompassing a diverse array of assets such as Ether, Matic, USDC, and more. Lastly, the variable denoted as "Amount USD" quantifies the transaction's monetary value in US dollars, affording a tangible measure of each transaction conducted on the Uniswap platform. This comprehensive dataset of transaction data forms the foundation for our empirical analysis, enabling us to explore and interpret the intricate dynamics of the DeFi landscape.

Figure~\ref{fig:workflow} delineates the intricate workflow of our data engineering process, serving as the foundation upon which our empirical analysis is built. Commencing with the acquisition of transaction data from Uniswap, we meticulously gather data spanning four distinct blockchain networks: Ethereum, Optimism, Arbitrum, and Polygon. To ensure the utmost rigor in our data collection, we perform robustness checks, which involve re-downloading all transaction data across different time points and conducting thorough comparisons to confirm consistency and completeness. This process helps identify and rectify any discrepancies or potential errors in the dataset, ensuring reliability for subsequent analysis.

The raw dataset procured through this process encompasses a diverse array of categories, including crucial elements such as timestamp information and transaction amounts. Through a series of transformative steps, we refine this raw data into a comprehensive output, which encompasses a multitude of essential metrics. These metrics encompass date-related information, along with statistical measures such as minimum, maximum, sum, standard deviation, variance, mean, median, 25th percentile, and 75th percentile. Furthermore, our dataset provides a detailed breakdown of transaction amounts in US dollars, along with the computation of four distinct decentralization indices.

This expansive dataset represents a trove of invaluable information, underpinning our research endeavors to unravel the intricate dynamics of the DeFi landscape. With over 50 million transactions recorded, originating from both Layer 1 (L1) and Layer 2 (L2) networks, our repository of daily indices casts a revealing light on critical facets of the network, encompassing adoption trends, scalability assessments, decentralization metrics, wealth distribution profiles, and other pivotal dimensions within the multifaceted DeFi ecosystem.

Our focus on Uniswap data is a strategic choice, as Uniswap is the largest and most active decentralized exchange (DEX) by trading volume and liquidity, making it a comprehensive source of transaction data across both Layer 1 (L1) and Layer 2 (L2) networks. Its dominance in the decentralized finance (DeFi) space offers unparalleled insights into economic shifts driven by the adoption of L2 solutions. While there are other DEXs such as SushiSwap and Curve, they either have smaller market shares or do not provide the same level of integration across multiple L2 networks, which is crucial for our analysis.
By leveraging Uniswap’s rich dataset, this study provides a nuanced understanding of the impact of L2 solutions on the DeFi economy, contributing valuable insights into network adoption, scalability, decentralization, and wealth distribution. In doing so, our research aims to enhance the collective understanding of this evolving ecosystem, offering a foundation for further research and innovations in decentralized finance.

%% file: 2.methods.tex
\section*{Methods}
Within this section, we delve into a comprehensive exposition of the intricacies underpinning our data engineering workflow, as visually depicted in Figure~\ref{fig:workflow}. Our goal is to provide a transparent and thorough account of the various stages and processes involved in the acquisition, transformation, and refinement of the extensive dataset that underlies our data processing. 

\subsection*{Data Collection}
In this study, we mainly rely on two data queries obtained from (\url{https://www.uniswap.shippooor.xyz/}.)

\begin{enumerate} 
    \item \textbf{Transaction Data set}: This detailed data set forms the basis of our main research inquiries.
    \item \textbf{Uniswap V3 Day Data set}: we use this data set to cross-verify our transaction data. 
\end{enumerate}

A data dictionary in Table~\ref{tab: data_dictionary} for these data sets can be found in Appendix~\ref{sec: additional_figures}. The Transaction data includes all successfully executed transactions (i.e., recorded on-chain) on the Uniswap V3 protocol across both the L1 network (Ethereum) and the L2 networks: Optimism, Arbitrum, and Polygon. We aggregate individual transaction data and compare it to the Uniswap V3 Day Data set. This comparison ensures data integrity and accuracy.
As shown in Figure~\ref{fig:gantt}
The L1 data spans from June 1, 2021, to December 31st, 2022, while the L2 data covers the period from the earliest initial launch date of the network pools, to December 31st, 2022. Polygon L2 starts from December 20, 2021, Arbitrum L2 starts from August 31 and Optimism L2 starts from November 11, 2022.
In this research, we primarily focus on the variable \textit{Amount USD}, which represents the transaction amount measured in US dollars for each transaction on Uniswap, since this will be instrumental in addressing our research questions. 

We have recorded a total of 24,720,474 transactions for the L1 network (Ethereum), amounting to a traded value of \$ 859,884,715,811.05, and 58,241,898 transactions for the L2 network (31,526,506 for Polygon, 17,428,305 for Optimism, 9,287,087 for Arbitrum), summing up to a total traded value of \$ 63,739,776,068 (\$ 34,045,289,633, \$ 6,982,218,140, and \$ 22,712,268,295 respectively).

\subsection*{Data Processing}
To assess decentralization within the DeFi ecosystem, we compute four key decentralization indices summarized in the paper "SoK: Blockchain Decentralization"~\cite{zhang2022sok}: Shannon Entropy, Herfindahl-Hirschman Index (HHI), Gini Coefficient, and Nakamoto Coefficient. These indices are the most commonly used measures in blockchain research, providing essential insights into the distribution and concentration of transaction values. Each metric captures a different aspect of decentralization—ranging from the overall diversity of participants (Shannon Entropy), the dominance of large players (HHI), the inequality in resource distribution (Gini Coefficient), to the threshold of control by a few entities (Nakamoto Coefficient).

\subsubsection*{Shannon Entropy (Entropy)}
Shannon Entropy measures the degree of disorder or uncertainty in a distribution. In our context, it quantifies the evenness of transaction values across users or addresses. The formula for calculating Shannon Entropy is as follows:

\begin{equation}
E = -\sum_{i=1}^{n} p_i \cdot \log_2(p_i)
\end{equation}

Where:
\begin{align*}
E & : \text{Shannon Entropy} \\
n & : \text{Number of users or addresses} \\
p_i & : \text{Proportion of total transaction value held by user or address } i
\end{align*}

This index captures the degree of concentration within the ecosystem, with higher values indicating greater decentralization.

\subsubsection*{Herfindahl-Hirschman Index (HHI)}
The HHI measures the concentration of transaction values among users or addresses. It is calculated as follows:

\begin{equation}
HHI = \sum_{i=1}^{n} s_i^2
\end{equation}

Where:
\begin{align*}
HHI & : \text{Herfindahl-Hirschman Index} \\
n & : \text{Number of users or addresses} \\
s_i & : \text{Proportion of total transaction value held by user or address } i
\end{align*}

The HHI assigns higher values for more concentrated distributions, reflecting reduced decentralization.

\subsubsection*{Gini Coefficient (Gini)}
The Gini Coefficient measures income or wealth inequality within a distribution. In our context, it assesses the inequality in transaction values among users or addresses. The formula for calculating the Gini Coefficient is as follows:

\begin{equation}
Gini = \frac{2 \cdot \sum_{i=1}^{n} i \cdot s_i}{n \cdot \sum_{i=1}^{n} s_i} - \frac{n + 1}{n}
\end{equation}

Where:
\begin{align*}
Gini & : \text{Gini Coefficient} \\
n & : \text{Number of users or addresses} \\
s_i & : \text{Proportion of total transaction value held by user or address } i
\end{align*}

The Gini Coefficient ranges from 0 to 1, with higher values indicating greater inequality.

\subsubsection*{Nakamoto Coefficient (Nakamoto)}
The Nakamoto Coefficient quantifies the proportion of total transaction value controlled by a specific percentage of the top users or addresses. It is calculated by sorting users or addresses by transaction value and summing their values until a cumulative threshold of 50\% is reached. The Nakamoto Coefficient reflects the concentration of transaction value among a smaller group of users or addresses.

Our decentralization index calculations are executed on a daily basis, providing the ability to monitor decentralization trends dynamically. Researchers and users can leverage the flexibility of our open-source Python code, designed using object-oriented programming principles, within Jupyter Notebook, to calculate indices for various time ranges. They also have the option to customize the frequency of index calculations to align with their specific research requirements. Additionally, we have incorporated the $tqdm$ package (\url{https://tqdm.github.io/}) into our codebase, enhancing user experience by enabling them to precisely track the progress of the code. While our default setting generates indices on a daily basis, users possess the flexibility to adjust the code to produce indices at shorter or longer time intervals, ensuring that the analysis aligns with the granularity required for their research objectives. Figure ~\ref{fig:entropy} in Section~\ref{sec: additional_figures} in the Appendix provides a visual representation of the entropy output for each of the four networks, offering insights into decentralization dynamics. Additional figures for the remaining indexes can be found in Appendix~\ref{sec: additional_figures}. Specifically, Figure ~\ref{fig:nakamoto} illustrates the Nakamoto index across different the four networks, Figure ~\ref{fig:HHI} presents the Herfindahl-Hirschman Index (HHI) by the four networks, and Figure ~\ref{fig:Gini} showcases the Gini index across the fvariours networks.


In addition to the decentralization indices, our Python code generates daily transaction volume statistics, encompassing a range of metrics such as minimum, maximum, sum, standard deviation, variance, mean, median, 25th percentile, 75th percentile, and transaction amounts in US dollars. Figure ~\ref{fig:transaction} visualizes the daily transaction trends across the four networks. Please note that the left axis labeled "count" represents the number of transactions on any given day, while the right axis labeled "sum" indicates the total US dollar value of all transactions within a given day. Furthermore, Figure ~\ref{fig:volume} in Appendix~\ref{sec: additional_figures} provides insights into daily trading volume by network, illustrating metrics such as mean, median, and 25th and 75th percentiles.


%% file: 3.datarecord_validation.tex
\section*{Data Records}
\label{sec:data records}
The final data records are stored and published on the Harvard Dataverse.\cite{DVN/OSOR3P_2023} The \texttt{raw.zip} file comprises the extracted data as delineated in the provided documentation, encapsulating the Transaction Dataset, Uniswap Day Dataset, and Liquidity Pool Dataset. This compilation includes diverse elements of blockchain transactions such as transaction hashes, timestamps, token specifics, and amounts in different currencies, along with daily aggregated statistics like volume and transaction count. The data originates from distinguished blockchain networks including Ethereum, Polygon, Arbitrum, and Optimism, and spans specific date ranges. Additionally, there are files containing calculated daily indices derived from the \texttt{raw.zip} data. These indices incorporate decentralization measures such as Entropy, Gini coefficient, Herfindahl-Hirschman Index (HHI), and the Nakamoto coefficient, in conjunction with detailed transaction volume statistics presenting metrics like minimum, maximum, sum, standard deviation, mean, median, and quantiles. The filenames of these data files commence with network identifiers: \texttt{"Eth\_"} for Ethereum, \texttt{"Polygon\_"} for Polygon, \texttt{"Arb\_"} for Arbitrum, and \texttt{"Opt\_"} for Optimism, succeeded by the name of the index or statistic (e.g., \texttt{Eth\_Entropy.csv}, \texttt{Polygon\_gini.csv}, \texttt{Arb\_HHI.csv}, \texttt{Opt\_nakamoto.csv}, \texttt{Eth\_volume.csv}). These prefixes aid in the systematic organization and effortless identification of the files, ensuring that analysts and researchers can straightforwardly navigate and scrutinize the data pertinent to each blockchain network. The records consist of the decentralized indices and transaction volume in \texttt{csv} format as shown in Table~\ref{tab:datasets}.

\textbf{Contents of Data Files}: The dataset comprises multiple types of data files, each serving a specific analytical purpose:
\begin{itemize}
    \item \textbf{Decentralization Indices}: We provide decentralization metrics in files such as Entropy.csv, gini.csv, HHI.csv, and nakamoto.csv. These indices offer valuable insights into the decentralization dynamics within the DeFi ecosystem, enabling researchers to assess the degree of decentralization across different networks.
    \item \textbf{Transaction Volume Statistics}: In the file volume.csv, we present comprehensive transaction volume statistics. This includes essential information such as minimum, maximum, sum, standard deviation, variance, mean, median, 25th percentile, 75th percentile, and transaction count. These statistics provide a holistic view of transaction volume trends and distribution within the DeFi landscape, facilitating in-depth analysis and research.
\end{itemize}
These distinct data files empower researchers to explore and compare DeFi activities, adoption patterns, decentralization levels, wealth distribution, and transaction volumes across Ethereum, Polygon, Arbitrum, and Optimism. Detailed information on the contents and usage of these data files is available in the accompanying Appendix, enhancing the dataset's utility for empirical analysis and research endeavors.

\textbf{Data Coverage and Observations}: 
The dataset captures a substantial temporal span, aligning with the specific time frames of the four blockchain networks:
\begin{itemize}
    \item \textbf{Ethereum}: For the Ethereum network, our dataset covers the time frame from June 1, 2021, to December 31, 2022.
    \item \textbf{Polygon}: For the Polygon network, our dataset covers the time frame from 2021-12-20 to 2022-12-31, 2022. 
    \item \textbf{Arbitrum}: Data from the Arbitrum network spans from 2021-08-31 to 2022-12-31.
    \item \textbf{Optimism}: The Optimism network's data encompasses the period from 2021-11-11 to 2022-12-31 
\end{itemize}

These distinct time ranges ensure that researchers can explore and compare DeFi activities, adoption patterns, and network dynamics within Ethereum, Polygon, Arbitrum, and Optimism, offering a comprehensive view of the evolving DeFi landscape on these networks. It's essential to refer to these specific time frames for precise temporal context in the dataset. All timestamps in the dataset remain consistently recorded in Coordinated Universal Time (UTC$+$0), maintaining uniformity and precision for temporal references.

\section*{Technical Validation}
\label{sec: validation}

In addition to the transaction dataset, we have meticulously collected data from the Uniswap V3 daily dataset for both Layer 1 (L1) and Layer 2 (L2) networks. This dataset provides crucial insights into the total daily trading volume on Uniswap, represented in both Ethereum (ETH) and US dollars (USD).

We utilize this supplementary dataset to validate the accuracy of our transaction data. Specifically, we compare the \textit{volumeUSD} variable from the Uniswap daily dataset with the daily summation of the \textit{Amount USD} variable in our transaction dataset. Our analysis reveals that, for the L1 network, our transaction data comprehensively covers 99\% of the total trading volume over the 19-month period. Similarly, for the L2 networks, our transaction data captures more than 99\% of the total trading volume for each respective network and period.

Figure~\ref{fig:daily_validation} visually illustrates the percentage of traded volume captured by our transaction dataset in comparison to the volume recorded in the Uniswap daily dataset. Notably, for both L1 and L2 networks, the majority of days exhibit a 100\% match, with only a few instances showing minor deviations. Further scrutiny of the data confirmed that these minor discrepancies are not due to errors in the data querying process but are likely attributed to variations in the conversion of transaction sizes to USD for specific transactions. Importantly, these minor deviations are not expected to significantly impact the integrity of our primary analysis results.

%% file: 4.UsageNotes.tex
\section*{Usage Notes}
\subsection*{Applicability}
Our dataset serves as a valuable tool for diverse fields including economics, data science, computer science, engineering, finance, and social sciences. Its broad applicability enables researchers to explore a range of analytical and research questions across multiple disciplines as exemplified below. 

\begin{enumerate}
\item \textbf{Analyzing Blockchain Network Adoption and Evaluating Scalability Solutions}: The daily indices produced in this research provide an indispensable tool for the rigorous monitoring and in-depth analysis of daily trading activities across diverse blockchain networks. These indices are critical in illuminating key aspects such as network adoption, trends in popularity, and the temporal shifts in user engagement within these networks. A thorough examination of transaction volumes over time enables researchers to effectively assess the extent of network adoption, drawing upon theoretical frameworks and empirical findings presented in existing literature. \cite{akram2020adoption, cong2021tokenomics} The significance of our dataset is not confined to merely observing fluctuations within singular networks; rather, it facilitates an extensive, comparative analysis across different blockchain platforms. This comparative approach is vital for a holistic understanding of the blockchain ecosystem. Furthermore, the dataset is of paramount importance for researchers focused on scrutinizing the scalability of blockchain technologies. The scalability aspect is integral for the widespread application and adoption of blockchain, with potential far-reaching benefits for the broader community.\cite{zhou2020solutions} Through detailed examination of the transaction volumes within each network, over various time periods, researchers can glean critical insights into each network's capability to manage and efficiently process different magnitudes of transactional loads. This analysis is key to understanding and improving blockchain networks' throughput and scalability, which are essential components for their practical viability and future growth.

\item \textbf{Examining the Decentralization and Inequality in Blockchain Network Utilization}: Our dataset includes four distinct parameters that provide insight into the levels of decentralization and inequality across various blockchain networks. Despite the initial conceptualization of blockchain technology as a means to establish an economy founded on decentralized trust, recent scholarly inquiries have highlighted a significant discrepancy between the theoretical ideal of network decentralization and its practical implementation. Studies have indicated a tendency towards centralization within various layers of blockchain infrastructure and a somewhat myopic technological singularity perspective that often overlooks the critical aspect of decentralization in actual network utilization. \cite{zhang2022sok,Zhang23,ao2022decentralized,zhang2023network} The concept of 'decentralization in usage' pertains to the quantitative assessment of user contributions to economic activities within blockchain networks and the distribution of these activities among network participants. Researchers utilizing our dataset are poised to conduct a nuanced exploration of the contrasts between Layer 1 (L1) and Layer 2 (L2) networks in terms of this decentralization metric. This analysis is pivotal for evaluating the influence of L2 solutions on the economic dynamics of blockchain networks. Additionally, It is important to explore how decentralization in network usage interacts with key blockchain performance indicators, such as efficiency and security, as highlighted by recent studies\cite{fu2024quantifying}. Decentralization can enhance security by distributing control across more participants, reducing the risk of malicious attacks, but it may also introduce inefficiencies, such as slower transaction throughput and increased latency due to the complexity of consensus mechanisms. Our framework offers indices to quantify decentralization, which can be used to analyze these trade-offs and identify the optimal balance between decentralization, efficiency, and security. Additionally, the link between transaction decentralization and social media engagement, along with its impact on geospatial economic growth, is significant. For instance, Chen et al.\cite{chen2024global} analyzed over 150 million geotagged tweets from 150 countries, showing how public sentiment toward decentralized finance (DeFi) varies by region and economic status. Their findings suggest that more developed countries engage more in discussions, while less developed nations exhibit higher sentiment levels. These insights demonstrate the importance of considering spatiotemporal and economic factors when assessing the implications of decentralized systems for global economic growth.

\item \textbf{Designing Blockchain as the Web3 Infrastructure for An Intelligent World}: In the endeavor to establish Blockchain as the core infrastructure for Web3 in an intelligent world, the significance of the dataset presented in this research cannot be overstated. Comprising over 50 million transactions from both Layer 1 and Layer 2 networks, this extensive and open-source dataset is pivotal for understanding the evolving dynamics of Decentralized Finance (DeFi). It offers crucial insights into the advancement of network efficiency and scalability brought about by Layer 2 solutions, addressing a critical gap previously hindered by the scarcity of comprehensive transaction data. Crucially, the size and openness of this dataset are fundamental in harnessing the intelligence of advanced machine learning models, particularly deep learning. As highlighted in Zhang's (2023)~\cite{zhang2023machine} study at the NeurIPS 2023 AI for Science Workshop, the intersection of AI and blockchain is a key driver in the digital era. This dataset not only illuminates various network dynamics, such as adoption trends and decentralization, but also provides a robust foundation for applying sophisticated AI techniques to blockchain data. The integration of these large-scale, rich datasets with cutting-edge machine learning algorithms is essential for developing more nuanced, intelligent blockchain technologies, thereby facilitating the emergence of a more interconnected and intelligent digital ecosystem.
\end{enumerate}

\subsection*{Future Research Directions}
The scope for future research in this domain is extensive and offers numerous avenues for enhancing and broadening our current understanding, particularly with respect to our daily indices. These include:

\begin{enumerate}
\item \textbf{Expanding the Dataset with Additional Network Attributes}: While the current daily indices offer a robust foundation for empirical analysis within the DeFi ecosystem, it is evident that the breadth of our dataset can be broadened to include more comprehensive variables. These variables should encapsulate diverse facets of DeFi network systems, as referenced in existing literature,\cite{zhang2023design,zhang2023network} including advanced network features,\cite{ao2022decentralized} intricate transaction fee mechanisms, \cite{liu2022CCS,zhang2023understand} and the nuanced social sentiment dynamics \cite{Fu2023AIEthics,Xiao_2023,quan2023decoding} prevalent in DeFi networks. Network features from social and economic networks may not serve as direct measures of decentralization but can provide insights into the degrees of decentralization by implying how power or control is distributed across participants in a system. A particular area of interest for future investigations is the security dimension within these networks, as outlined in existing research.\cite{zhou2023sok} Incorporating data on aspects like smart contract vulnerabilities, network consensus mechanisms, and on-chain security protocols will enhance our comprehensive understanding of the DeFi ecosystems’ security and resilience.

\item \textbf{Developing Advanced Blockchain Transaction Indices}: The potential of our dataset extends to the creation of specialized blockchain transaction indices that transcend basic network features. This could involve constructing indices that track unique wallet address usage per network, analyze the availability and dynamics of liquidity pools, and quantify transaction volumes within specific blocks, as suggested in references.\cite{zhang2023design,yu2023bitcoin,XAI_Crpyto2023,liu2022deciphering,Pipeline2022,Zhang23} Such indices would offer invaluable insights into various dimensions of DeFi and FinTech applications and could act as pivotal indicators for cryptocurrency asset valuations.

\item \textbf{Integrating Diverse Networks and Federated Learning Approaches}: With the ongoing evolution of blockchain technologies, it becomes imperative to integrate a broader spectrum of networks, focusing on their unique usage patterns and characteristics. This includes extending the dataset to encompass additional decentralized exchange (DEX) protocols and blockchain layers beyond the realm of Uniswap\cite{yan2024analyzing,quan2023decoding,liu2024economics} and exploring cross-chain solutions beyond Ethereum, as indicated in existing studies.\cite{angeris2019analysis} Additionally, to optimize computational resources, future studies could explore the application of federated learning. This approach would allow for the generation of aggregate indices without the need for extensive data merging, relying instead on the shared learning models across diverse networks.\cite{Zhang23,fu2024quantifying}
\end{enumerate}

\subsubsection*{Call for interdisciplinary collaboration}
The interdisciplinary collaboration showcased in this research underscores the imperative to delve further into the complex and multifaceted nature of decentralized finance (DeFi) and blockchain technologies within the fintech sector. It is essential for experts from a variety of disciplines – encompassing economics, computer science and engineering, cryptography, finance, and beyond – to collaboratively investigate the intricate dynamics characterizing these systems.\cite{zhang2023design} This collective effort is crucial for spurring innovation and contributing significantly to the advancement of this rapidly developing field. The ongoing exploration of new variables and the fostering of interdisciplinary cooperation are pivotal in deepening our understanding of the DeFi landscape. Such efforts are expected to lead to substantial improvements in security, scalability, and decentralization within these cutting-edge financial ecosystems.

%% file: 5.codeavailability.tex
\section*{Code Availability}

Our data and codes are publicly accessible on GitHub at the following URL (\url{https://github.com/sunshineluyao/uniswap}.) The code is implemented in Python and presented within Jupyter Notebook files. The initial (latest) release, designated as \textcolor{blue}{v1.0.0} ({v1.1.1}), published on November 11, 2023 (December 5, 2023), is available under the MIT License (DOI: 10.5281/zenodo.10259603). Researchers, developers, and interested parties can access and utilize this open-source code repository to explore and analyze DeFi transaction data from Uniswap.

%% file: 6.acknowledgements.tex
\section*{Acknowledgements}
The authors express their gratitude to Saghar Mohseni, Victoria Zhiqing Zhang, Alexander Fischer, Jingtian Guo, Eric Zhang, Anirudh Parasrampuria, Bonny Wang, Shay Patil, Michael Jiang, and Weixiao Wang for their valuable research assistance in data collection. Zhang acknowledges support from the National Science Foundation China on the project entitled “Trust Mechanism Design on Blockchain: An Interdisciplinary Approach of Game Theory, Reinforcement Learning, and Human-AI Interactions (Grant No. 12201266).” Cong thanks Ripple's University Blockchain Research Initiative (UBRI), the Initiative of Cryptocurrencies \& Contracts, and the FinTech Initiative at Cornell for research funding.


\section*{Author contributions statement}

All authors have made significant contributions to this study, from conceptualization to the final manuscript. L.Z. led the methodology design, coded the data analysis programs, and developed the dataset, ensuring analytical robustness. N.C. and D.L. were pivotal in data collection, with N.C. also focusing on comprehensive documentation and D.L. on data validation, guaranteeing data integrity. E.J. was responsible for drafting the manuscript. W.C. oversaw the project, providing critical supervision and contributing to the manuscript's refinement.

\section*{Competing interests}

The author(s) declare no competing interests.

%% file: 7.figures_tables.tex
\begin{figure}[!htbp]
    \centering
    \includegraphics[width=.90\linewidth]{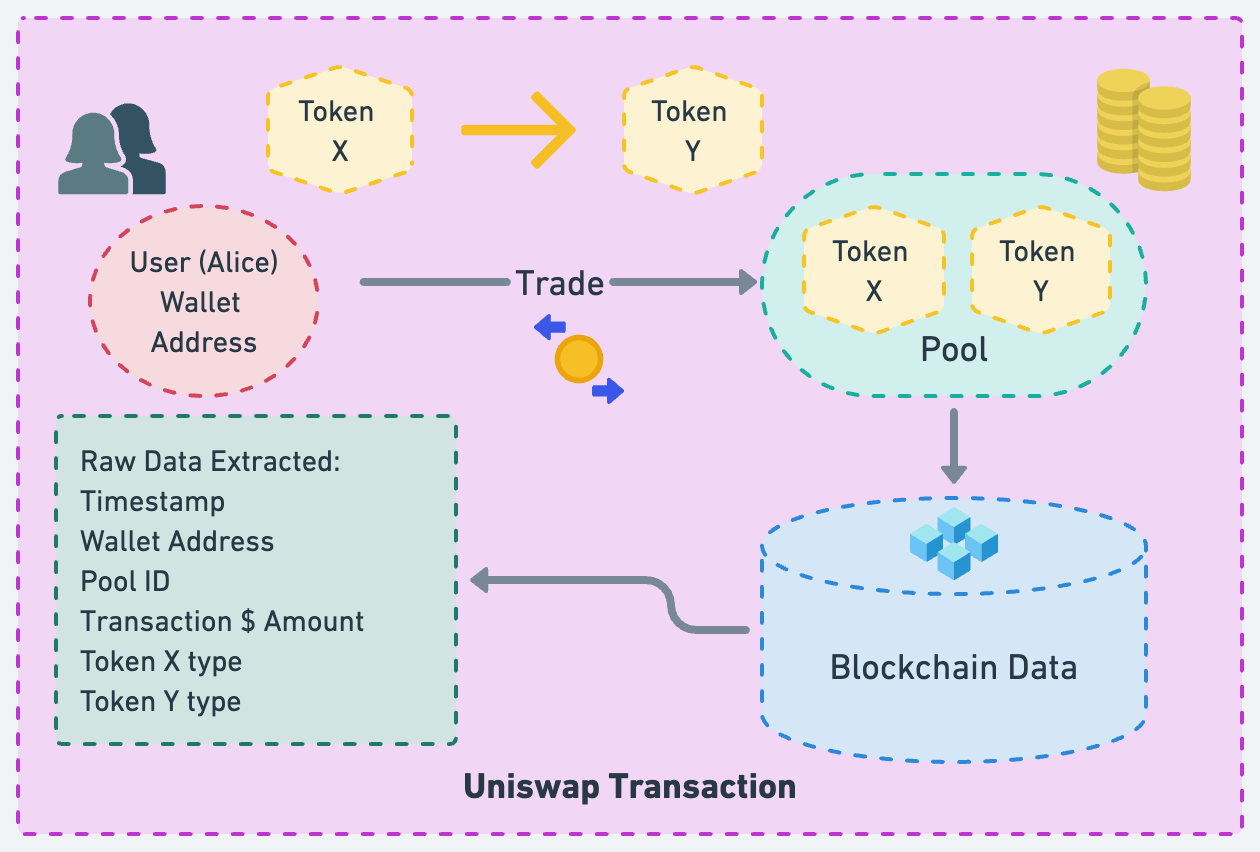}
    \caption{Transation Data on Uniswap}
    \label{fig:transaction_mindmap}
\end{figure}

\begin{figure}[!htbp]
    \centering
    \includegraphics[width=.8\linewidth]{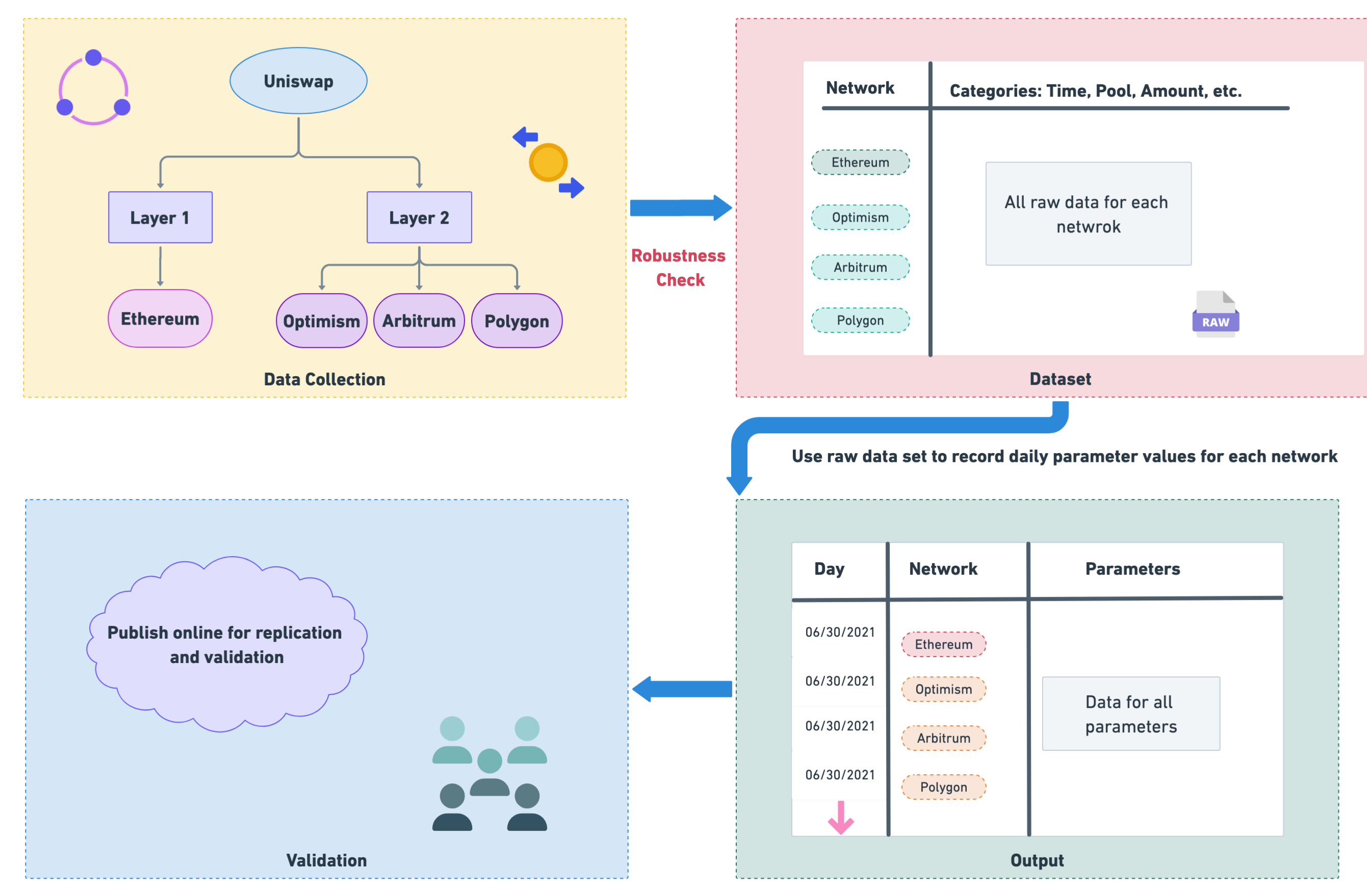}
    \caption{The Data Engineering Workflow. Start by collecting transactions trading data using Uniswap extractor tool at \url{https://www.uniswap.shippooor.xyz}}
    \label{fig:workflow}
\end{figure}

\begin{figure}
    \centering  
    \includegraphics[width=\linewidth]{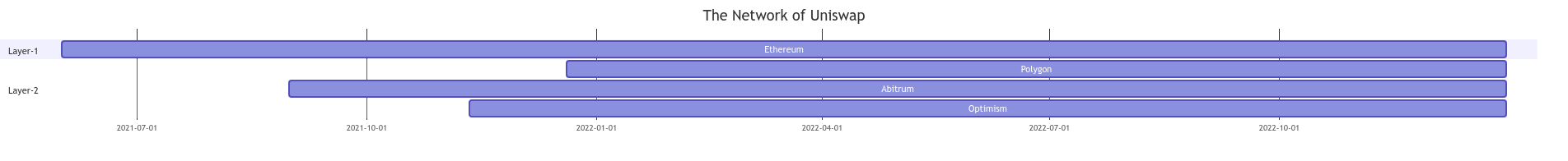}
    \caption{Data Range for the Networks of Uniswap.}
    \label{fig:gantt}
\end{figure}

\begin{figure}[hbtp!]

\begin{subfigure}{.475\linewidth}
  \includegraphics[width=\linewidth]{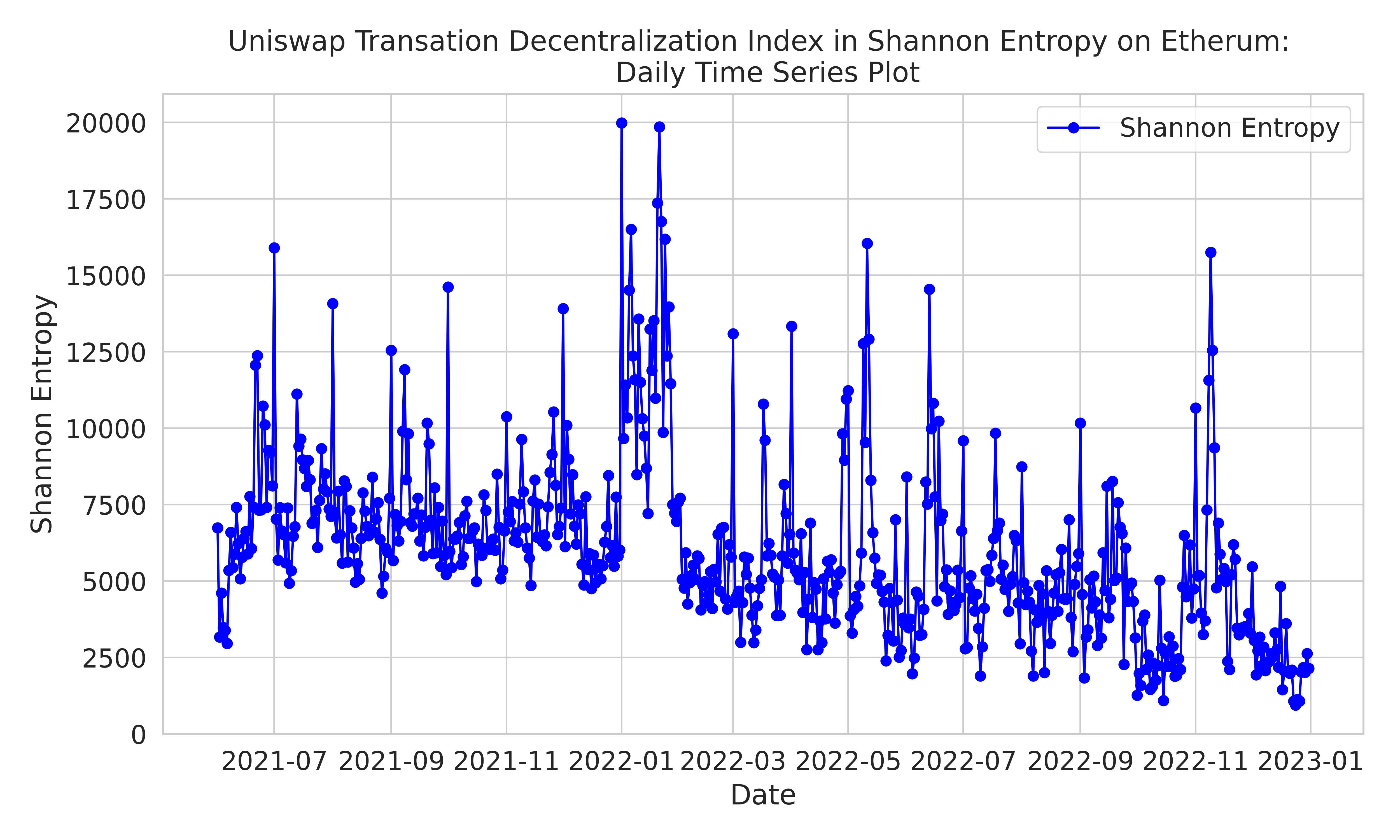}
  \caption{Entropy on Ethereum}
  \label{MLEDdet}
\end{subfigure}\hfill 
\begin{subfigure}{.475\linewidth}
  \includegraphics[width=\linewidth]{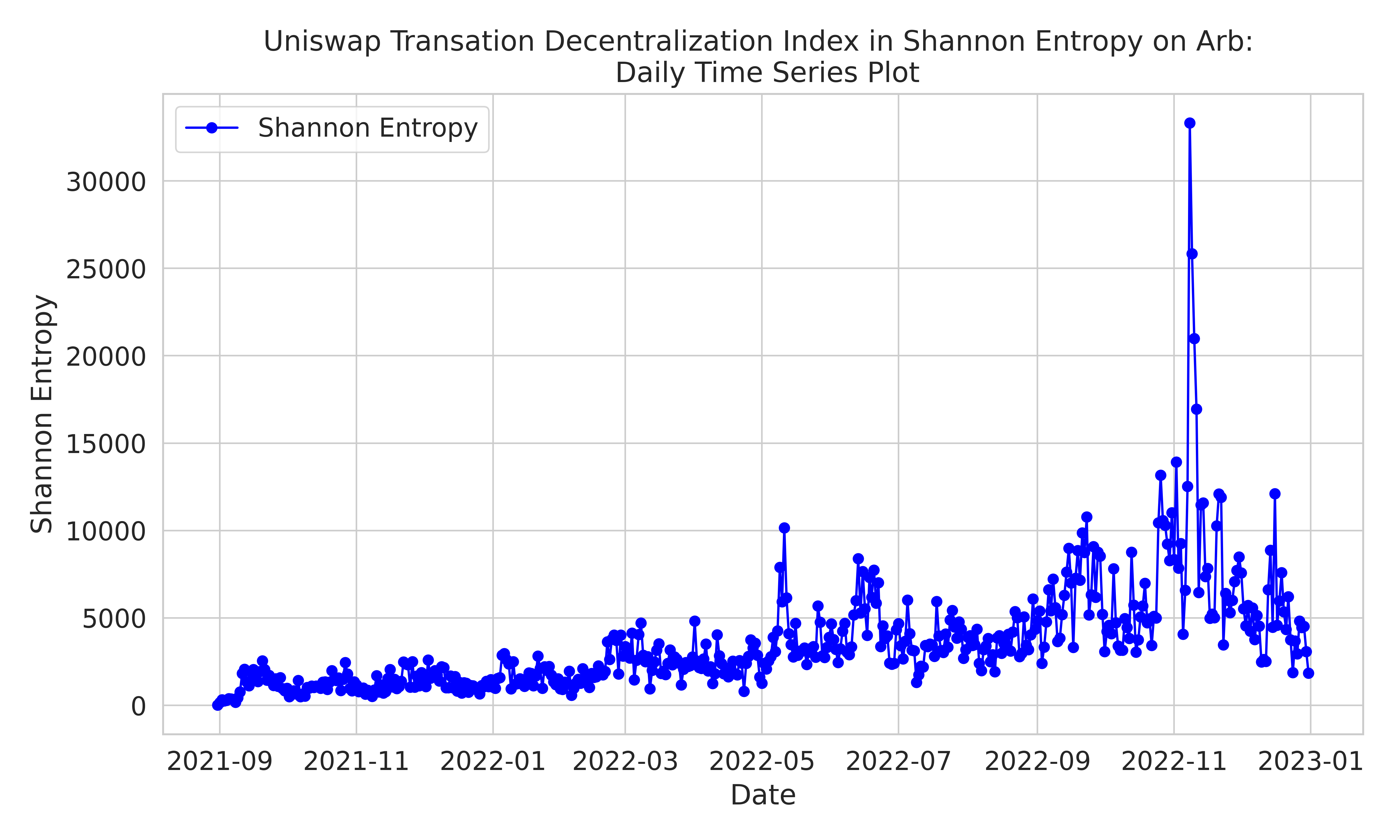}
  \caption{Entropy on Arbitrum}
  \label{energydetPSK}
\end{subfigure}

\medskip 
\begin{subfigure}{.475\linewidth}
  \includegraphics[width=\linewidth]{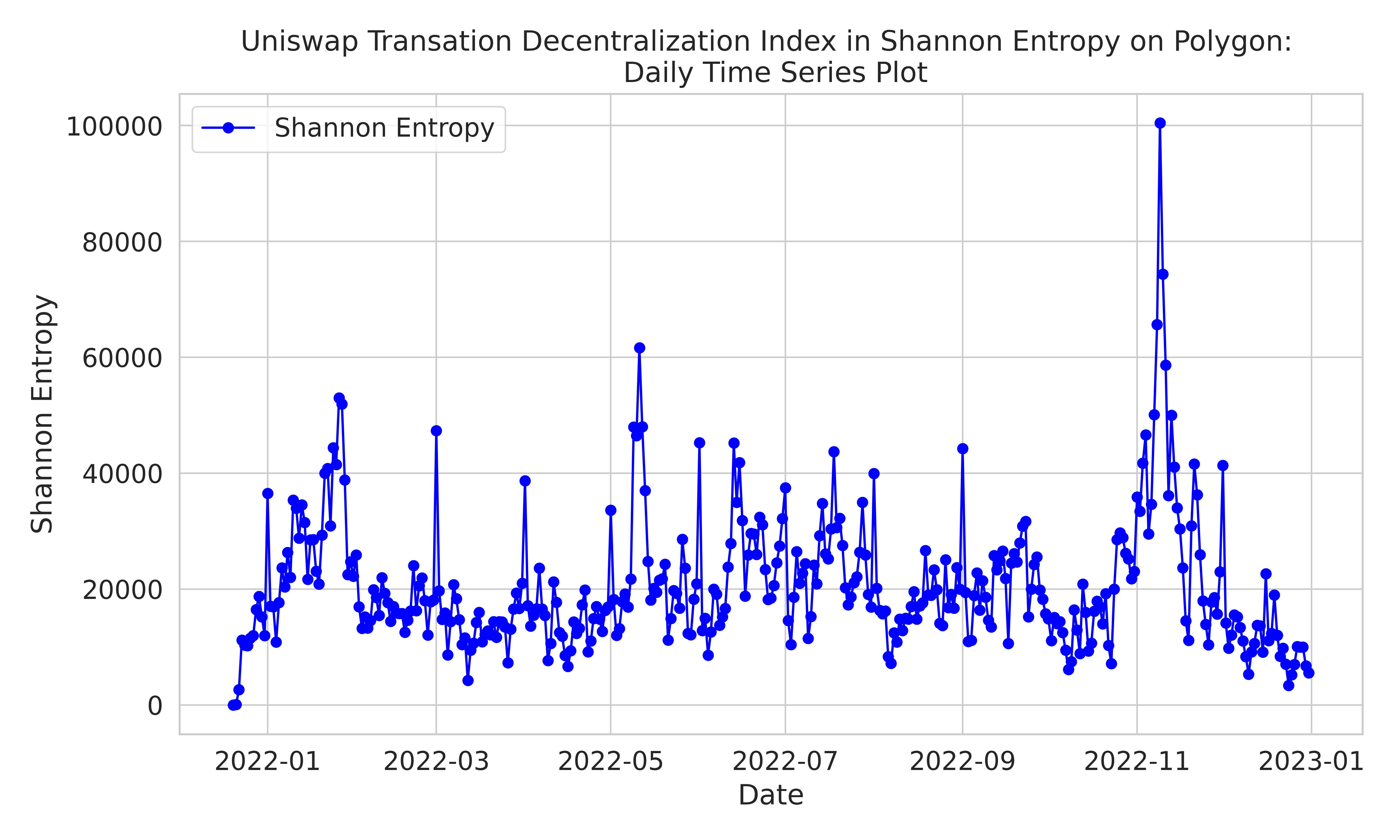}
  \caption{Entropy on Polygon}
  \label{velcomp}
\end{subfigure}\hfill 
\begin{subfigure}{.475\linewidth}
  \includegraphics[width=\linewidth]{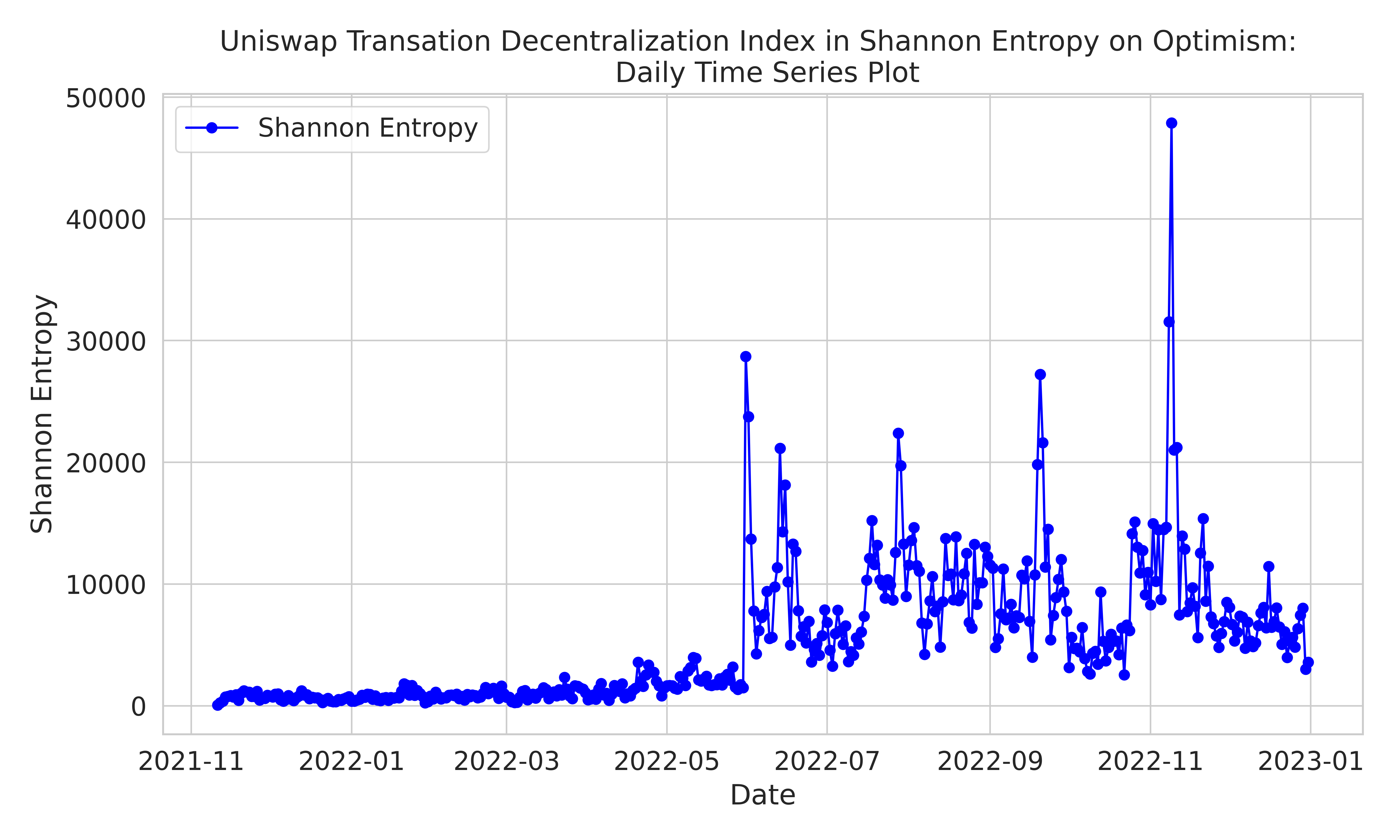}
  \caption{Entropy on Optimism }
  \label{estcomp}
\end{subfigure}

\caption{Daily Entropy Index by Network}
\label{fig:entropy}
\end{figure}

\begin{figure}[hbtp!]

\begin{subfigure}{.475\linewidth}
  \includegraphics[width=\linewidth]{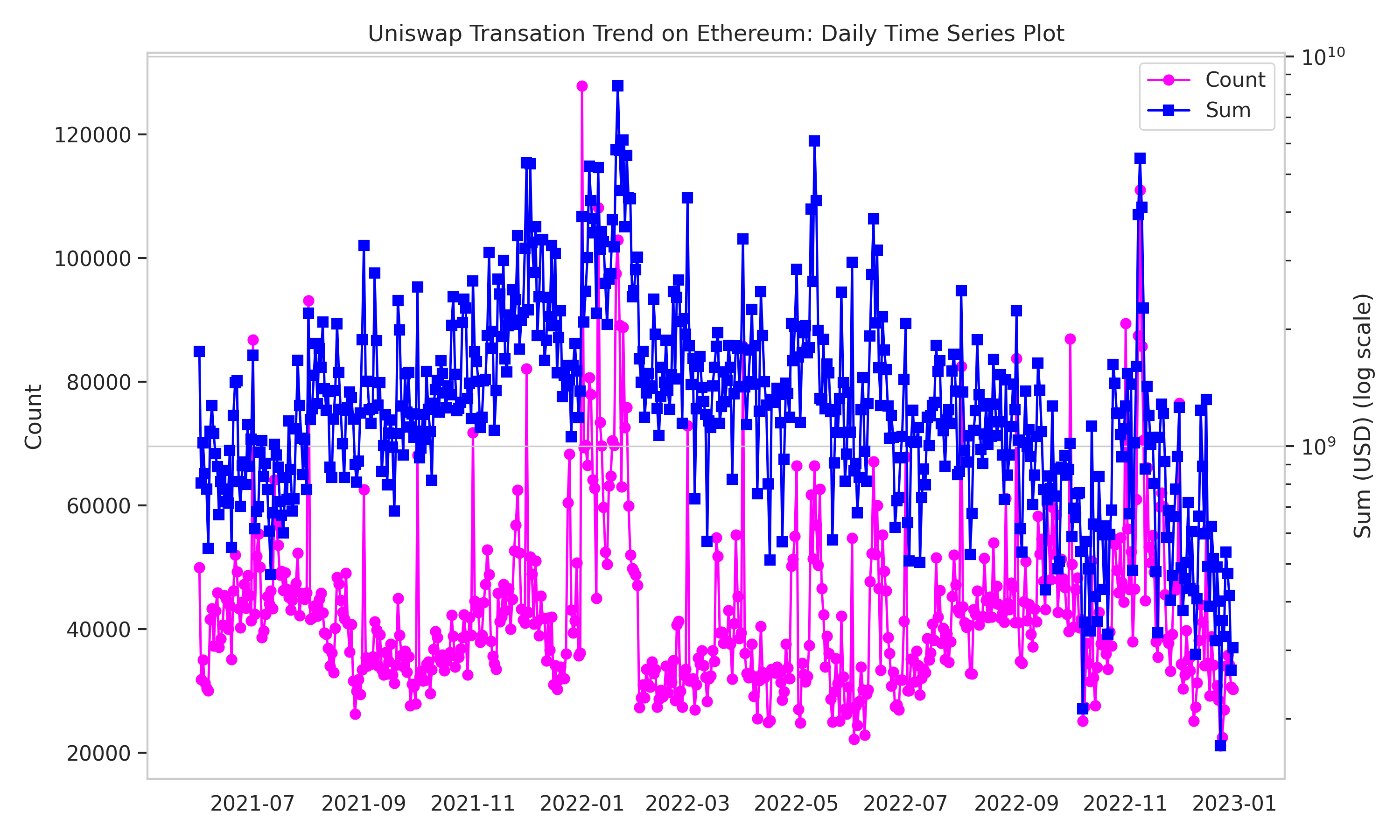}
  \caption{Daily Transaction on Ethereum}
  \label{MLEDdet}
\end{subfigure}\hfill 
\begin{subfigure}{.475\linewidth}
  \includegraphics[width=\linewidth]{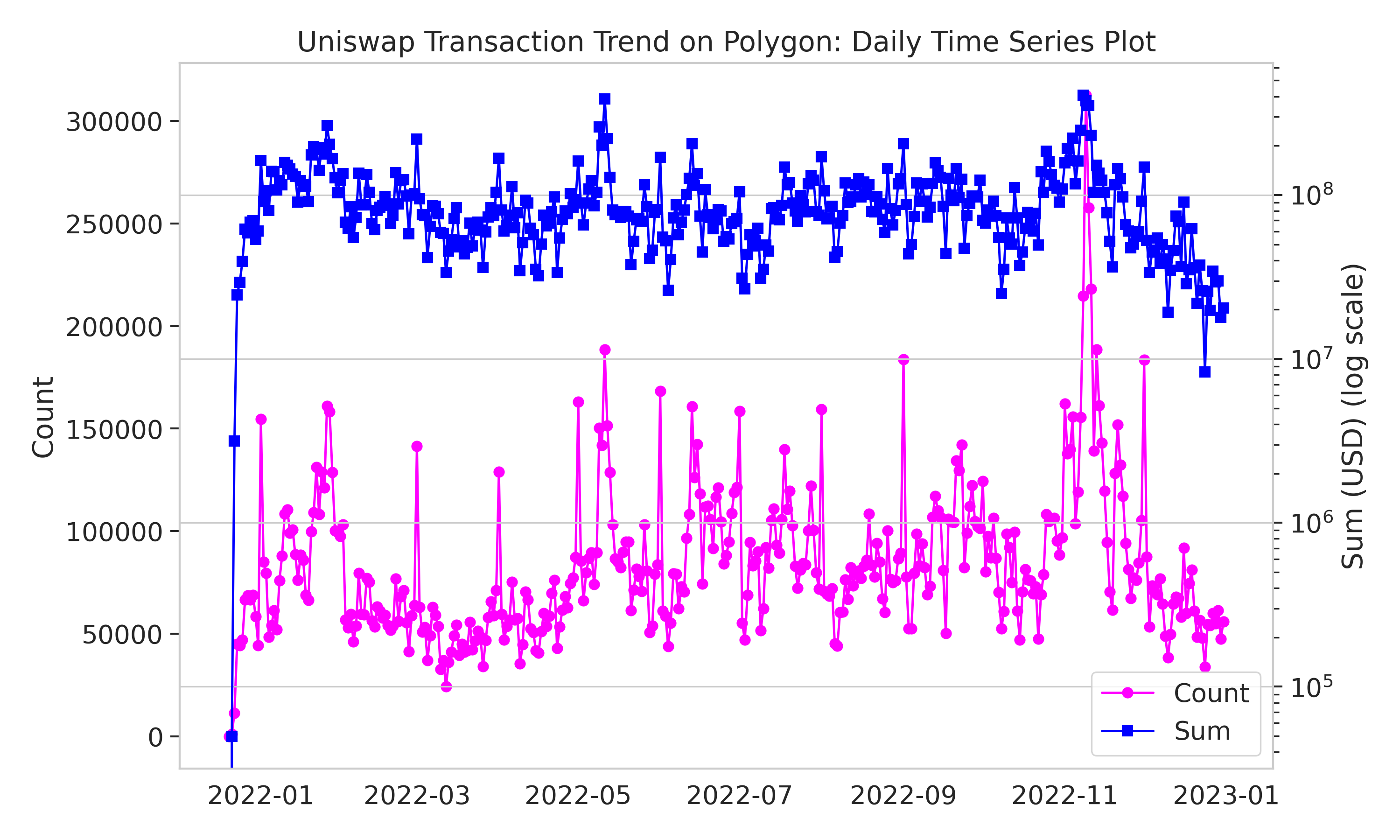}
  \caption{Daily Transaction on Polygon}
  \label{energydetPSK}
\end{subfigure}

\medskip 
\begin{subfigure}{.475\linewidth}
  \includegraphics[width=\linewidth]{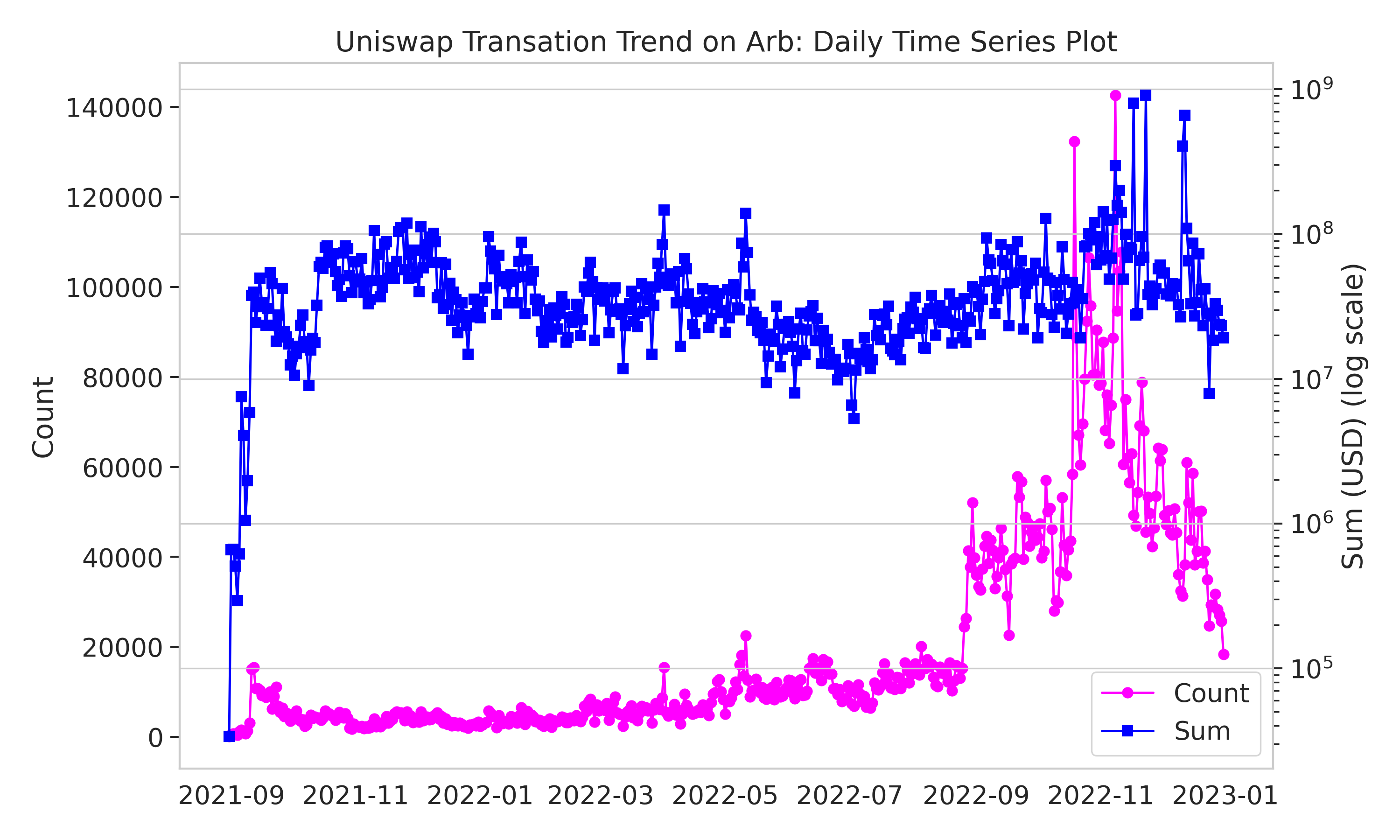}
  \caption{Daily Transaction on Arbitrum}
  \label{velcomp}
\end{subfigure}\hfill 
\begin{subfigure}{.475\linewidth}
  \includegraphics[width=\linewidth]{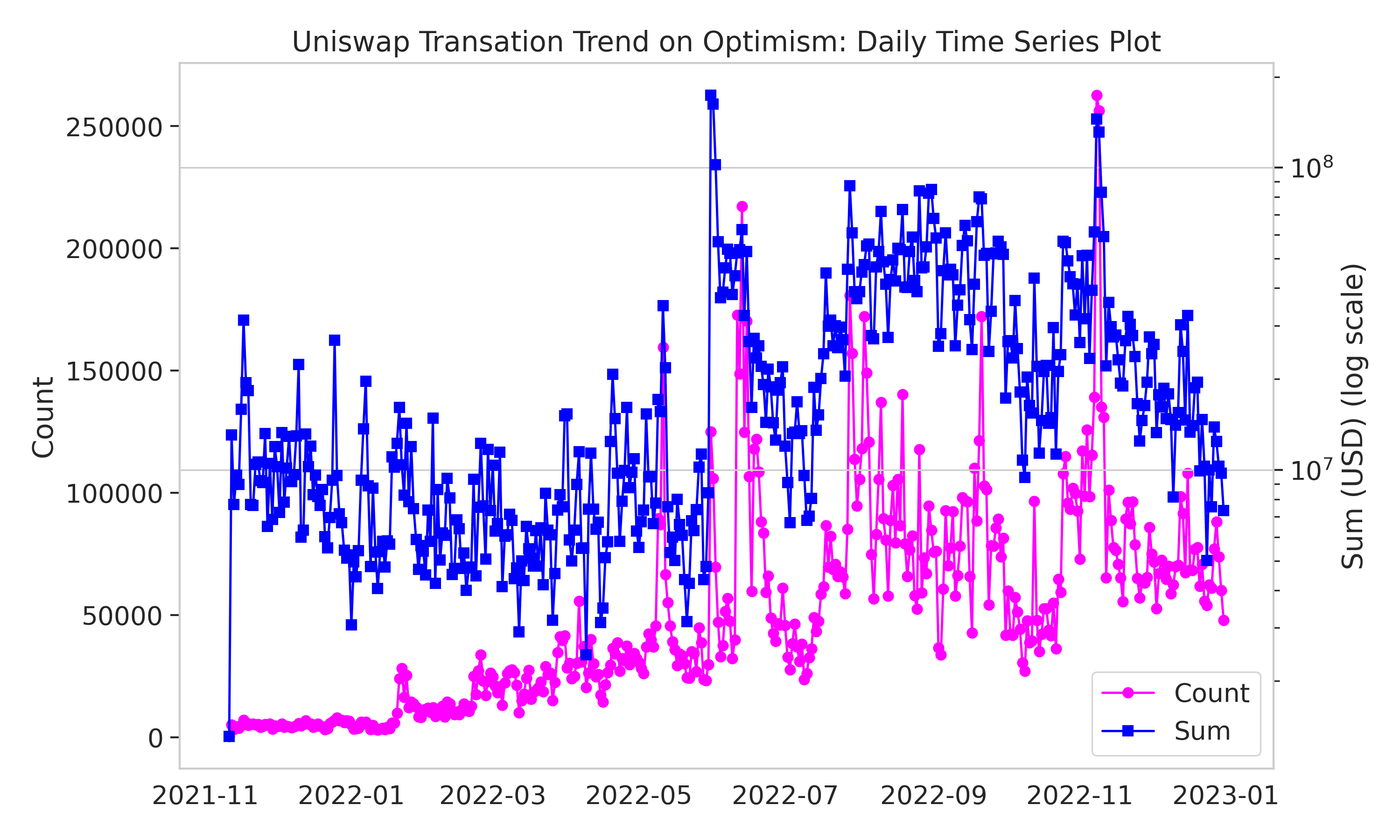}
  \caption{Daily Transaction on Optimism}
  \label{estcomp}
\end{subfigure}

\caption{Daily Transaction Trends by Network}
\label{fig:transaction}
\end{figure}

\begin{figure}[!hbtp]
\begin{subfigure}{.475\linewidth}
  \includegraphics[width=\linewidth]{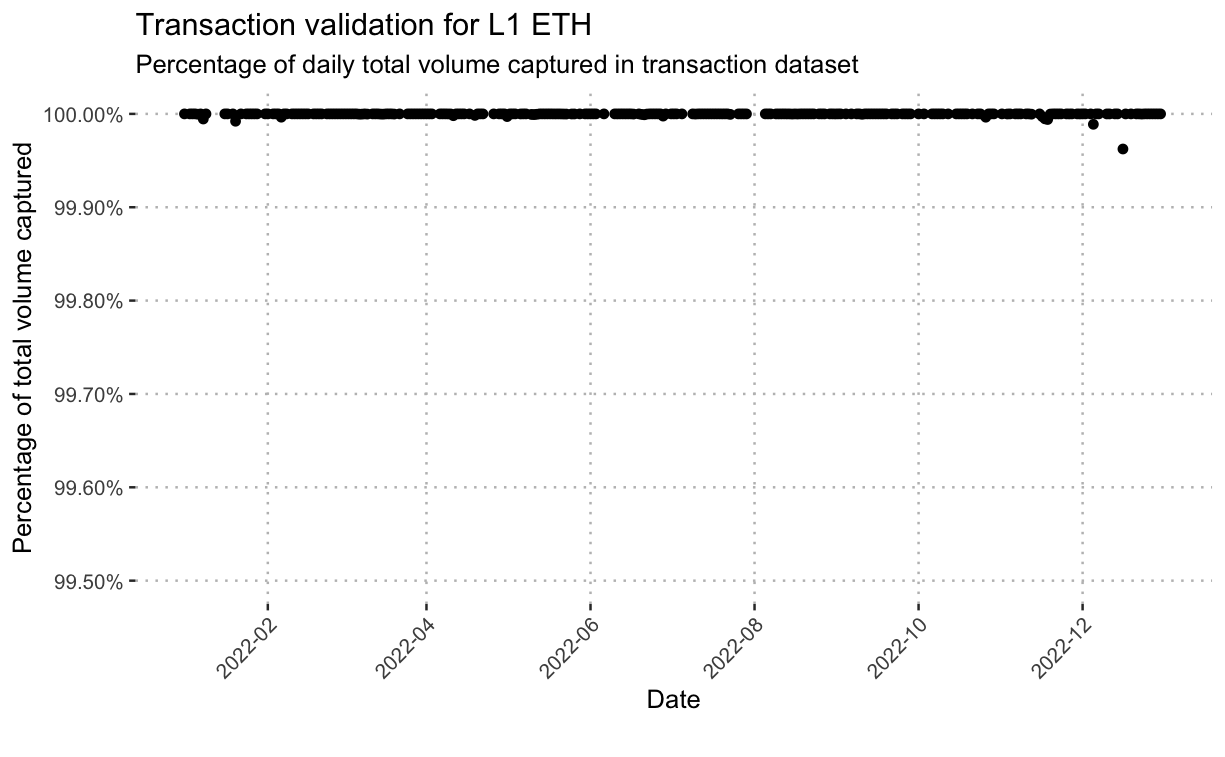}
  \caption{Ethereum}
  \label{MLEDdet}
\end{subfigure}\hfill 
\begin{subfigure}{.475\linewidth}
  \includegraphics[width=\linewidth]{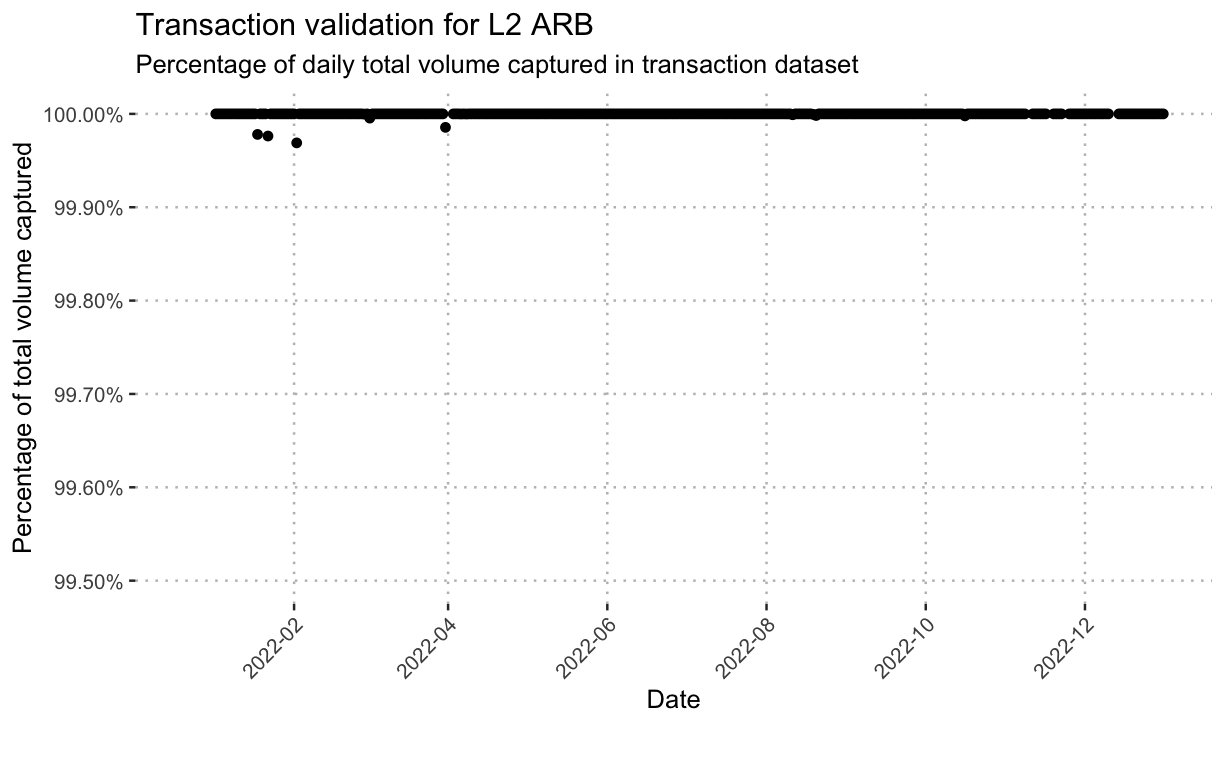}
  \caption{Arbitrum}
  \label{energydetPSK}
\end{subfigure}

\medskip 
\begin{subfigure}{.475\linewidth}
  \includegraphics[width=\linewidth]{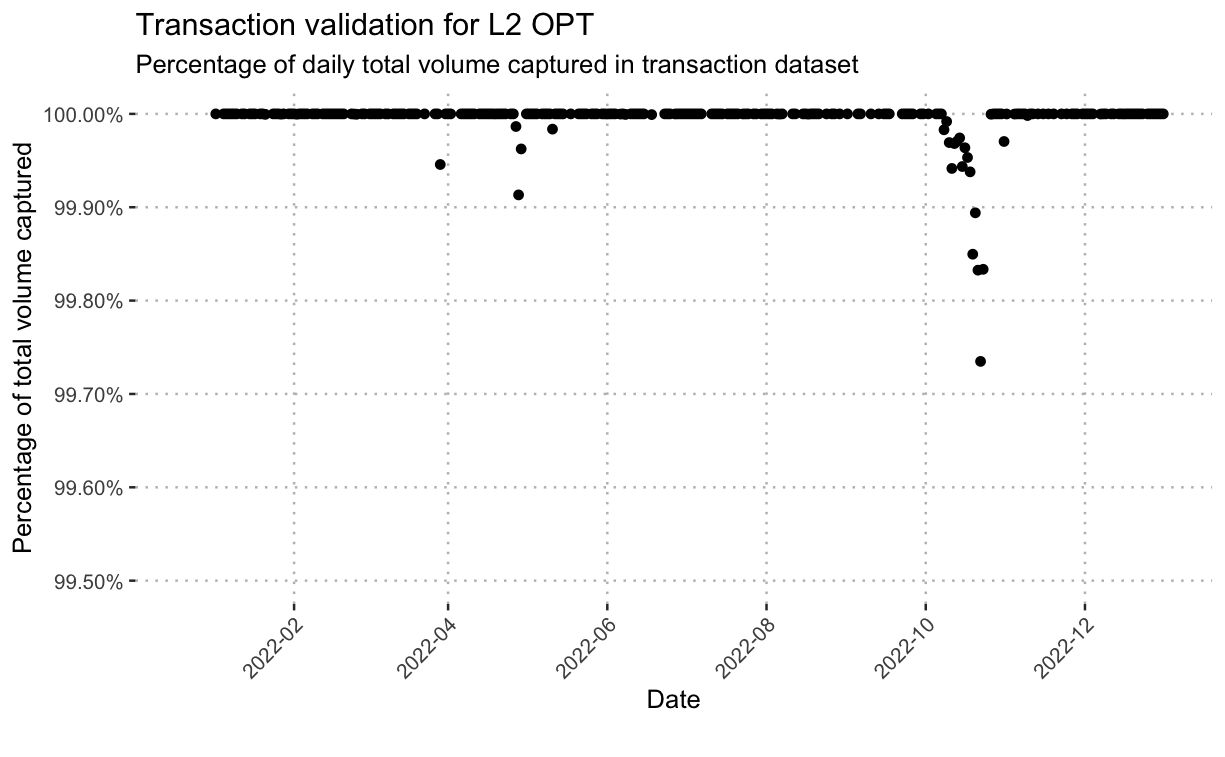}
  \caption{Optimisim}
  \label{velcomp}
\end{subfigure}\hfill 
\begin{subfigure}{.475\linewidth}
  \includegraphics[width=\linewidth]{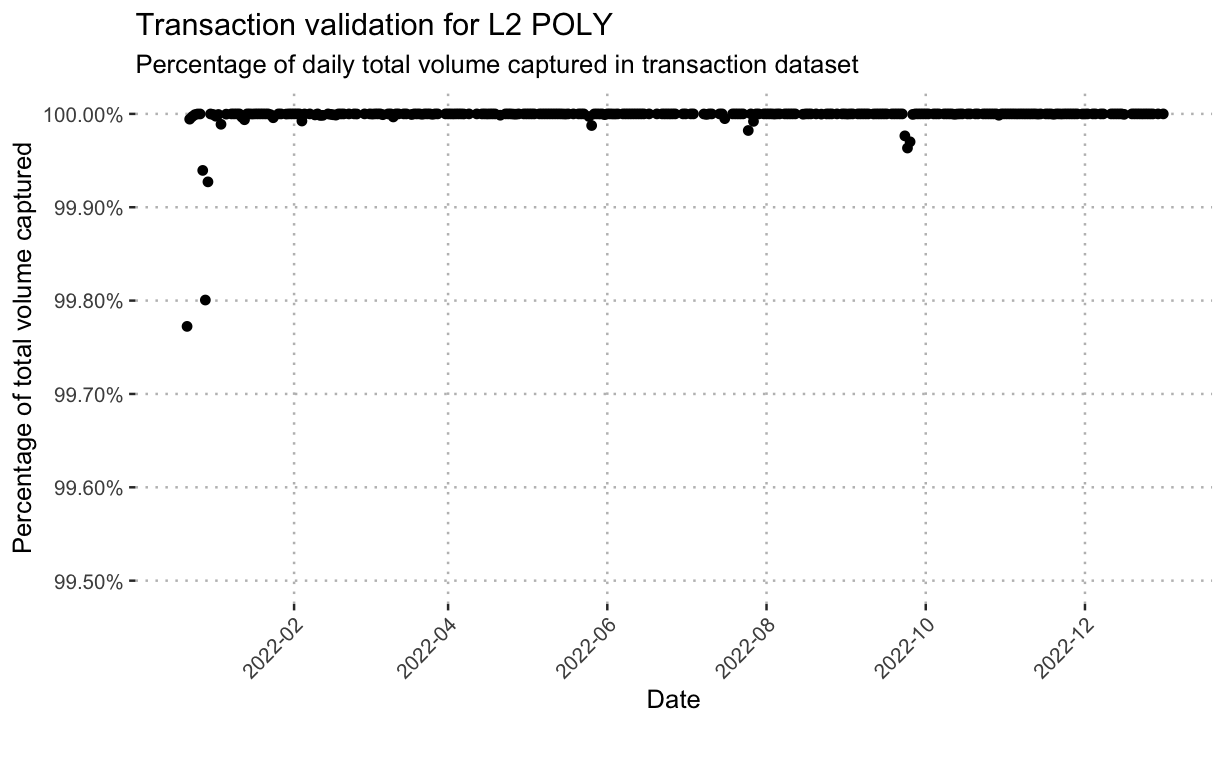}
  \caption{Polygon}
  \label{estcomp}
\end{subfigure}

\caption{Percentage of daily summation of traded volume captured in transaction data set compares to the daily total column in Uniswap day data set. Both measure in USD.}
\label{fig:daily_validation}
\end{figure}

\input{tabs/datasets}

%% file: tabs/datasets.tex
\begin{table}[!htbp]
\centering
\begin{tabular}{llll}
\toprule
\textbf{Network} & \textbf{Data Range (Start)} & \textbf{Data Range (End)} & \textbf{Data Files (Daily Frequency)} \\
\midrule
Ethereum & 2021-06-01 & 2022-12-31 & Decentralization Indices: Entropy.csv, gini.csv, HHI.csv, nakamoto.csv \\
Polygon & 2021-12-20 & 2022-12-31 & Transaction Volume Statistics: volume.csv \\
 & & & (min, max, sum, std, var, mean, median, 25\%, 75\% quantile, count) \\
Arbitrum & 2021-08-31 & 2022-12-31 & \\
Optimism & 2021-11-11 & 2022-12-31 & \\
\bottomrule
\end{tabular}
\caption{Data Records}
\label{tab:datasets}
\end{table}

%% file: 8.appendix.tex
\appendix


\clearpage

\section*{Additional Tables}
\label{sec: additional_tables}

\input{data_dictionary}

\clearpage

\section*{Additional Figures}
\label{sec: additional_figures}

\begin{figure}[hbt!]

\begin{subfigure}{.475\linewidth}
  \includegraphics[width=\linewidth]{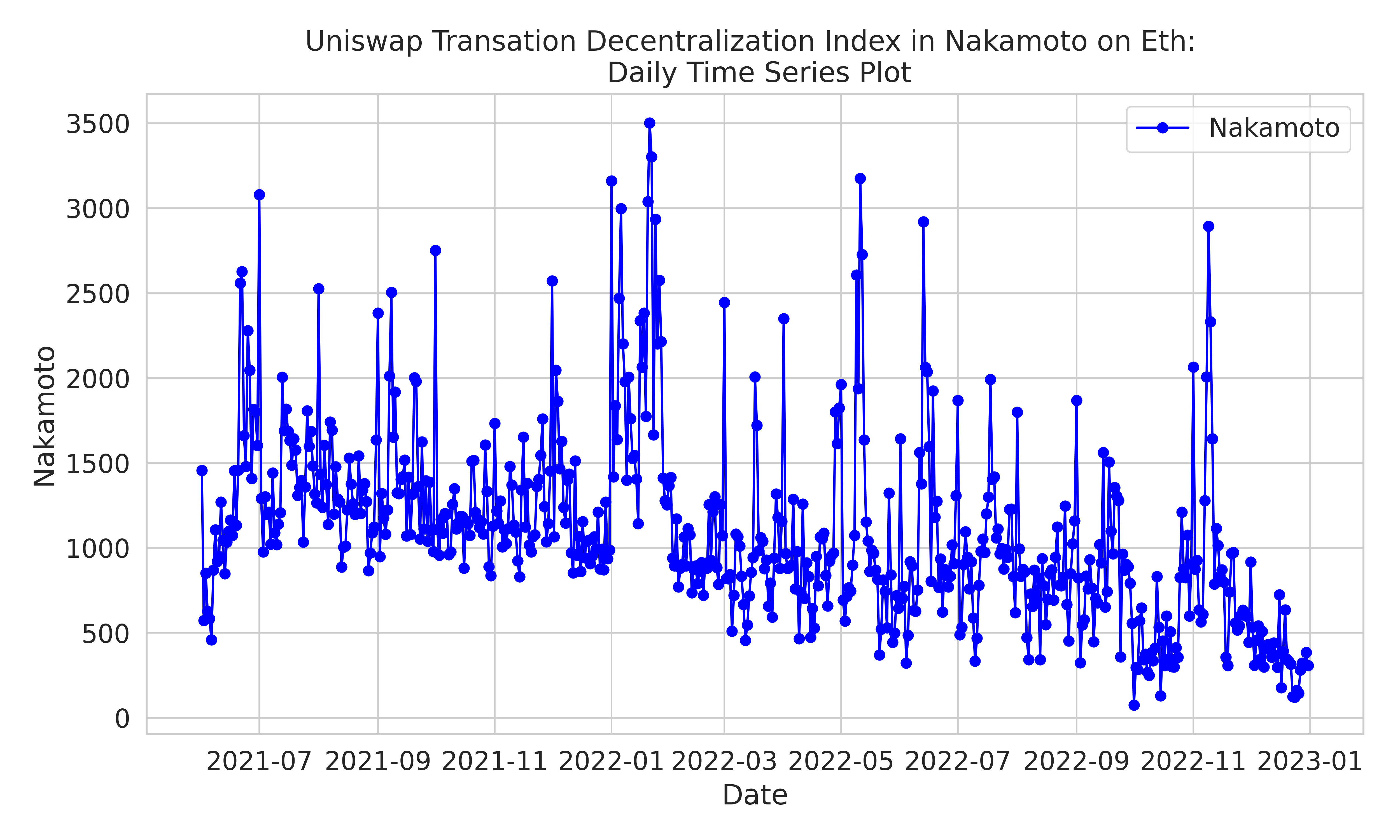}
  \caption{Nakamoto on Ethereum}
  \label{MLEDdet}
\end{subfigure}\hfill 
\begin{subfigure}{.475\linewidth}
  \includegraphics[width=\linewidth]{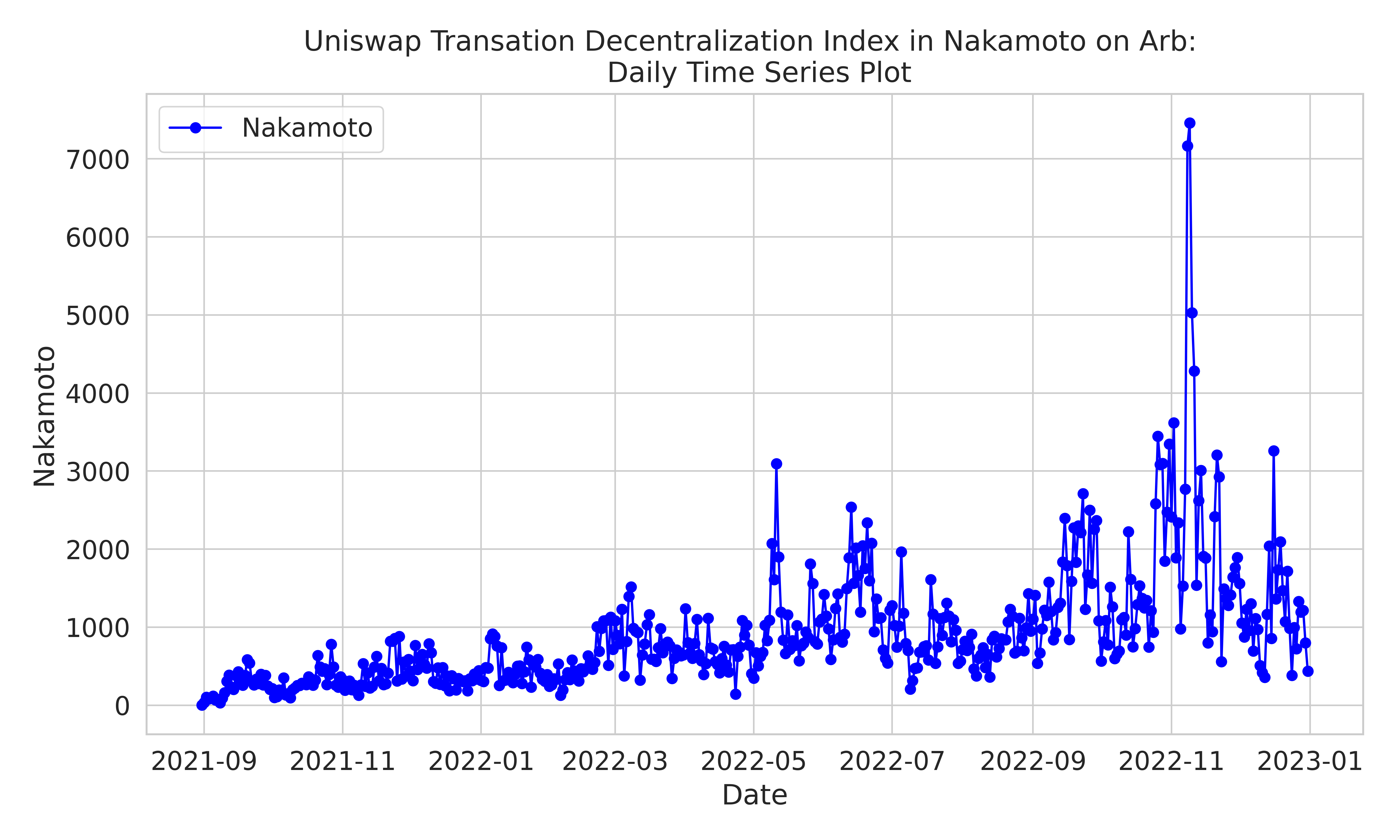}
  \caption{Nakamoto on Arbitrum}
  \label{energydetPSK}
\end{subfigure}

\medskip 
\begin{subfigure}{.475\linewidth}
  \includegraphics[width=\linewidth]{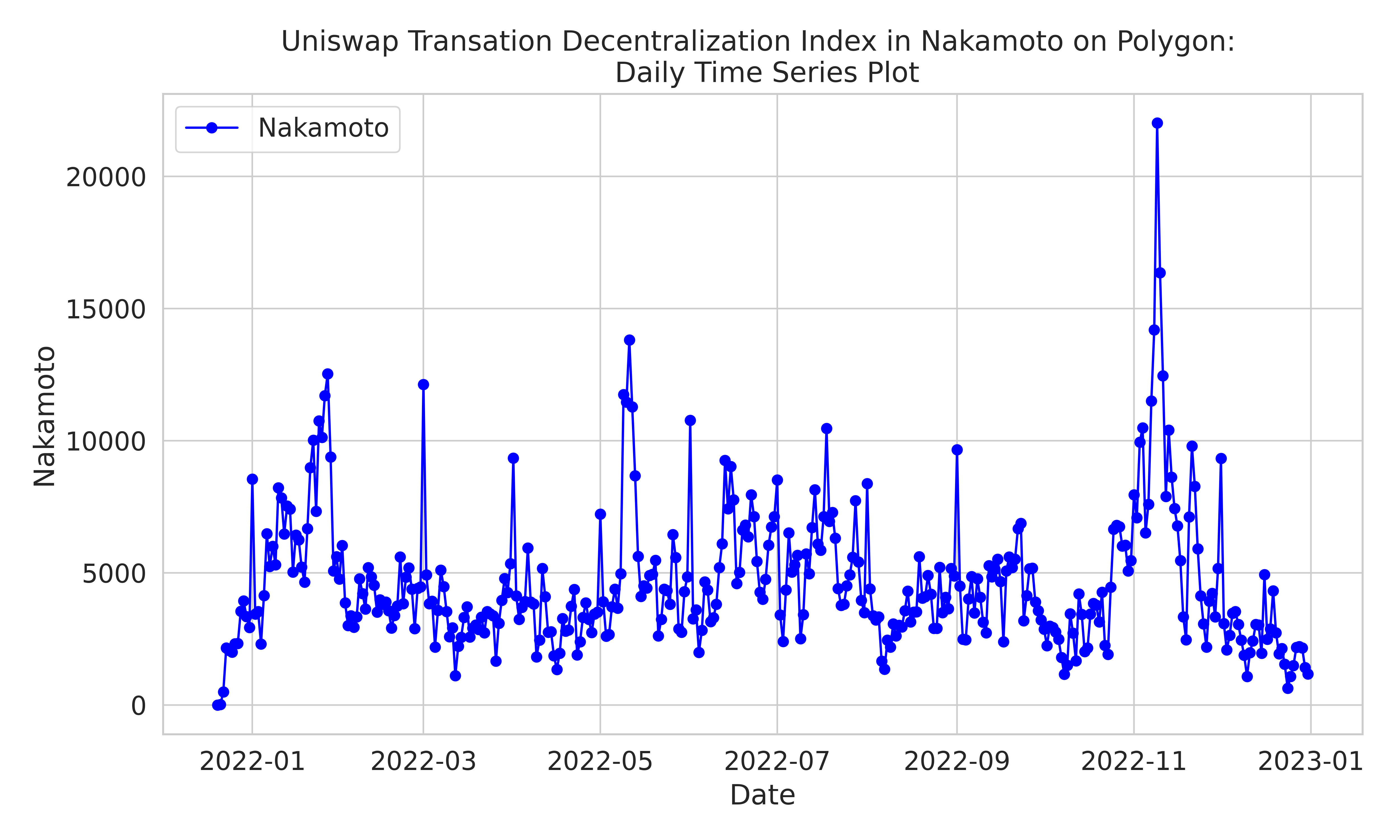}
  \caption{Nakamoto on Polygon}
  \label{velcomp}
\end{subfigure}\hfill 
\begin{subfigure}{.475\linewidth}
  \includegraphics[width=\linewidth]{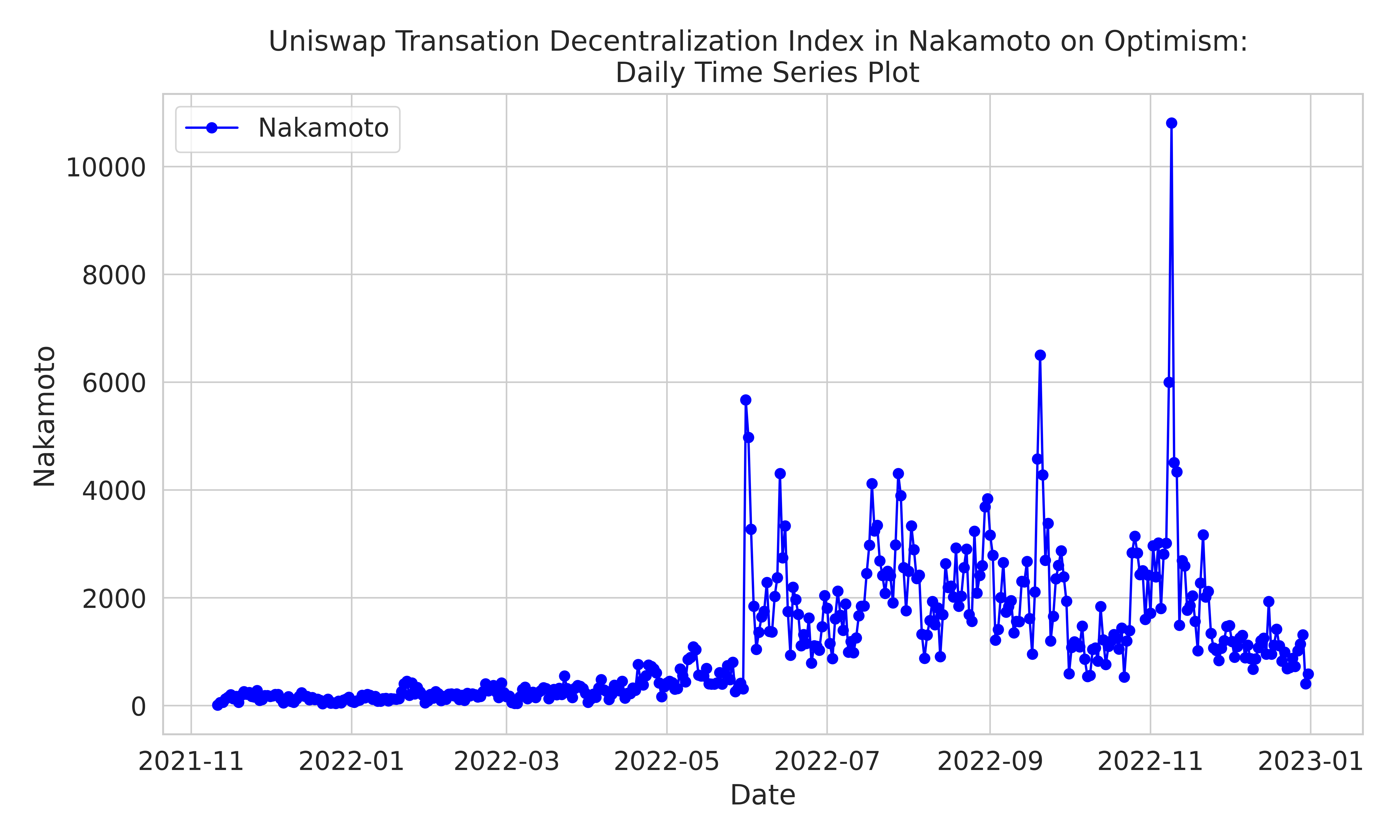}
  \caption{Nakamoto on Optimism }
  \label{estcomp}
\end{subfigure}

\caption{Daily Nakamoto Index by Network}
\label{fig:nakamoto}
\end{figure}

\begin{figure}[hbt!]

\begin{subfigure}{.475\linewidth}
  \includegraphics[width=\linewidth]{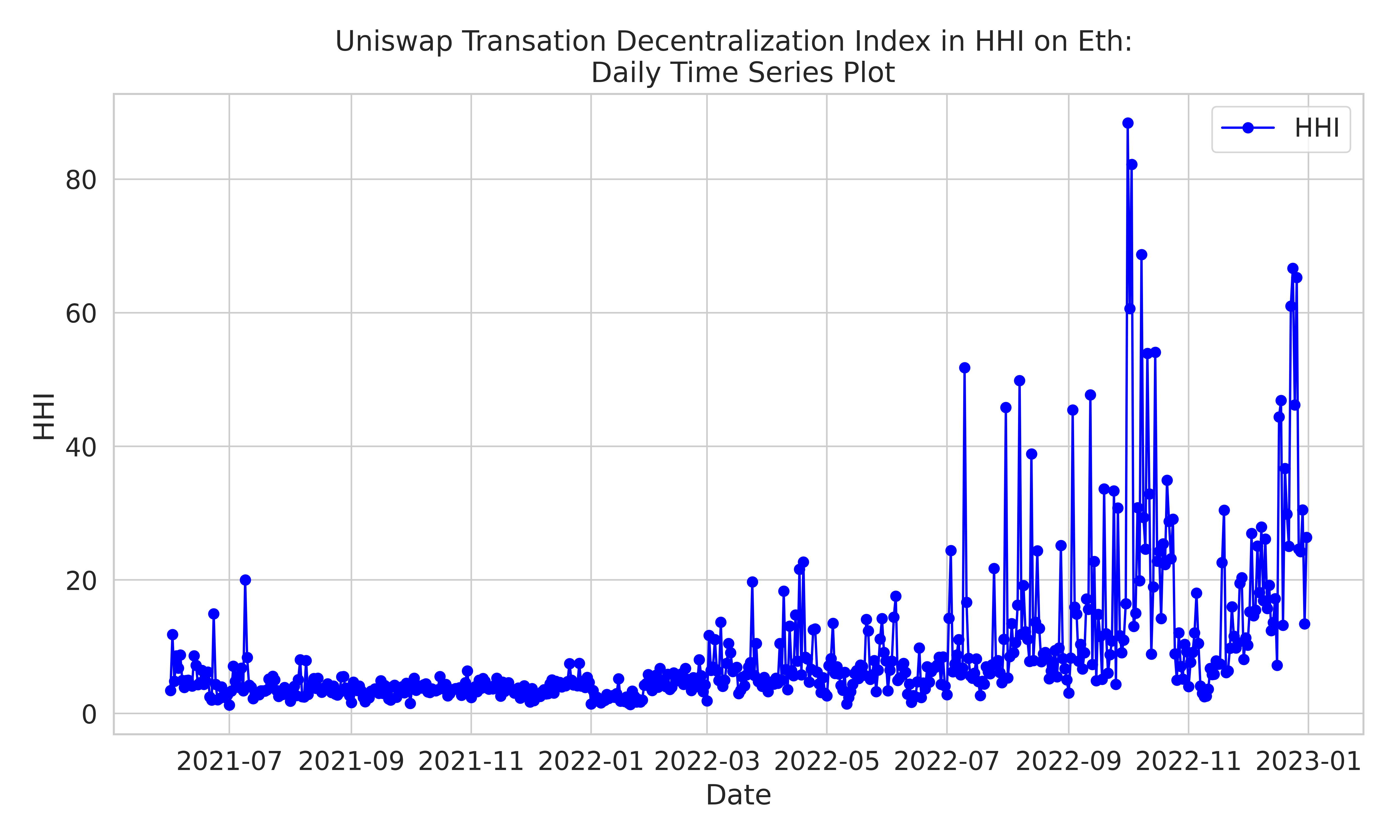}
  \caption{HHI on Ethereum}
  \label{MLEDdet}
\end{subfigure}\hfill 
\begin{subfigure}{.475\linewidth}
  \includegraphics[width=\linewidth]{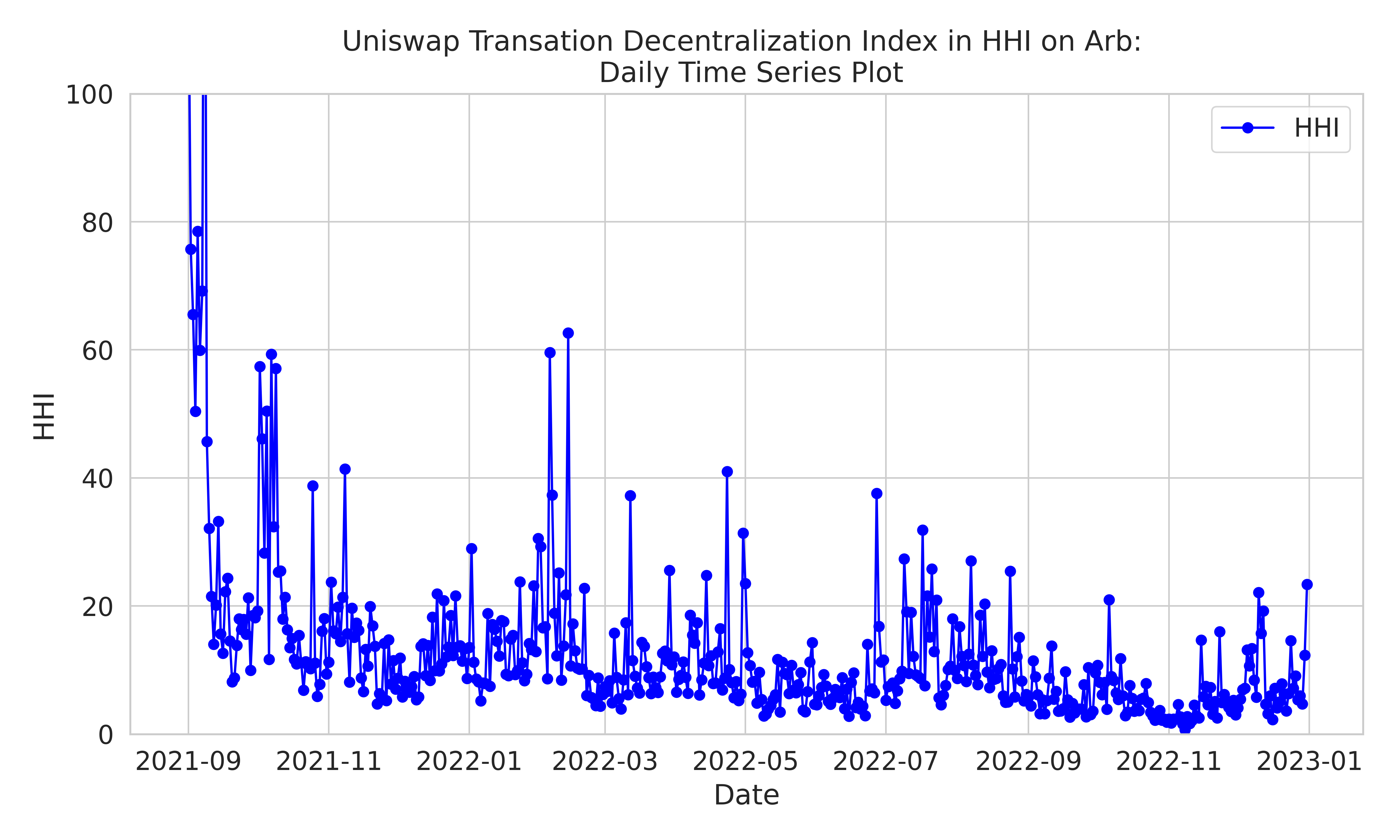}
  \caption{HHI on Arbitrum}
  \label{energydetPSK}
\end{subfigure}

\medskip 
\begin{subfigure}{.475\linewidth}
  \includegraphics[width=\linewidth]{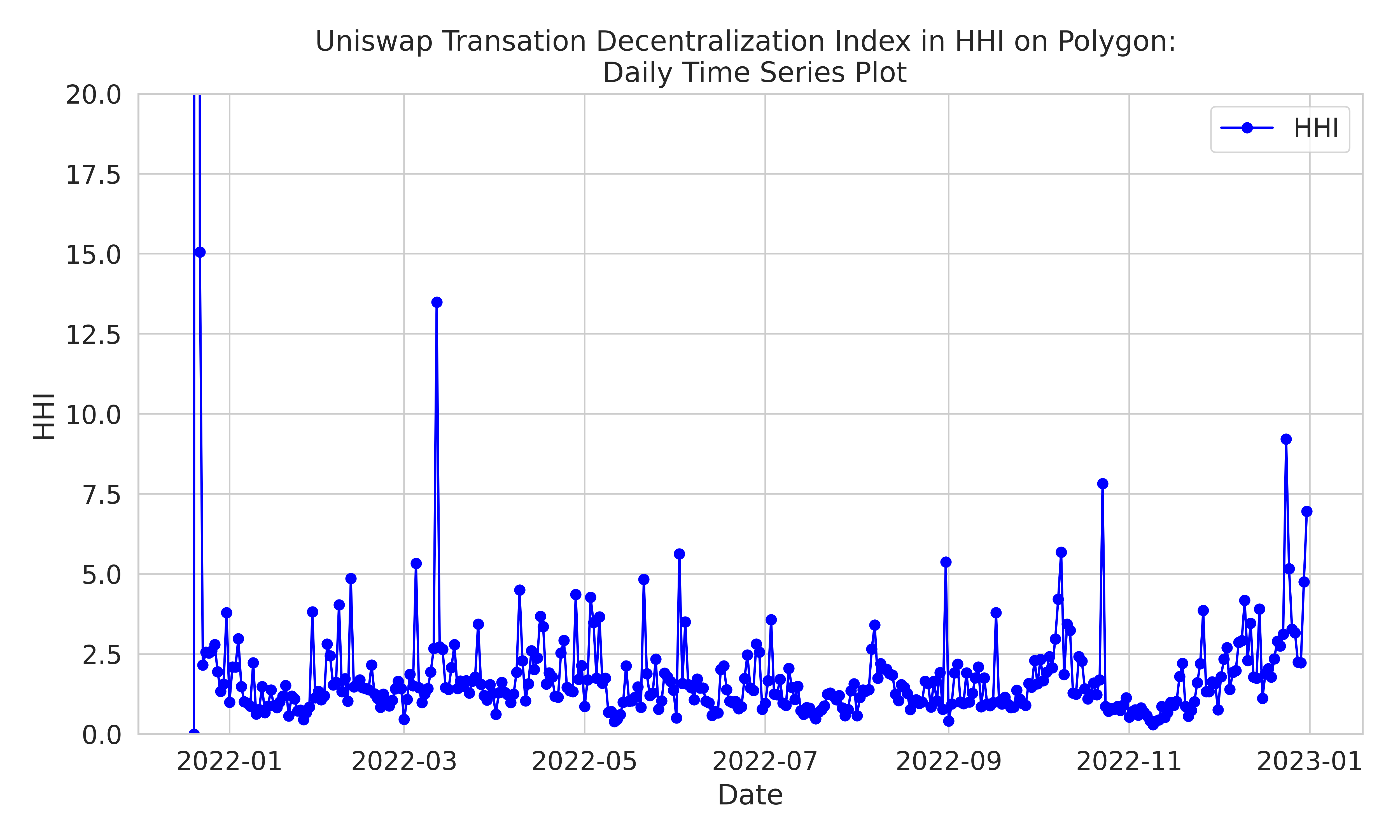}
  \caption{HHI on Polygon}
  \label{velcomp}
\end{subfigure}\hfill 
\begin{subfigure}{.475\linewidth}
  \includegraphics[width=\linewidth]{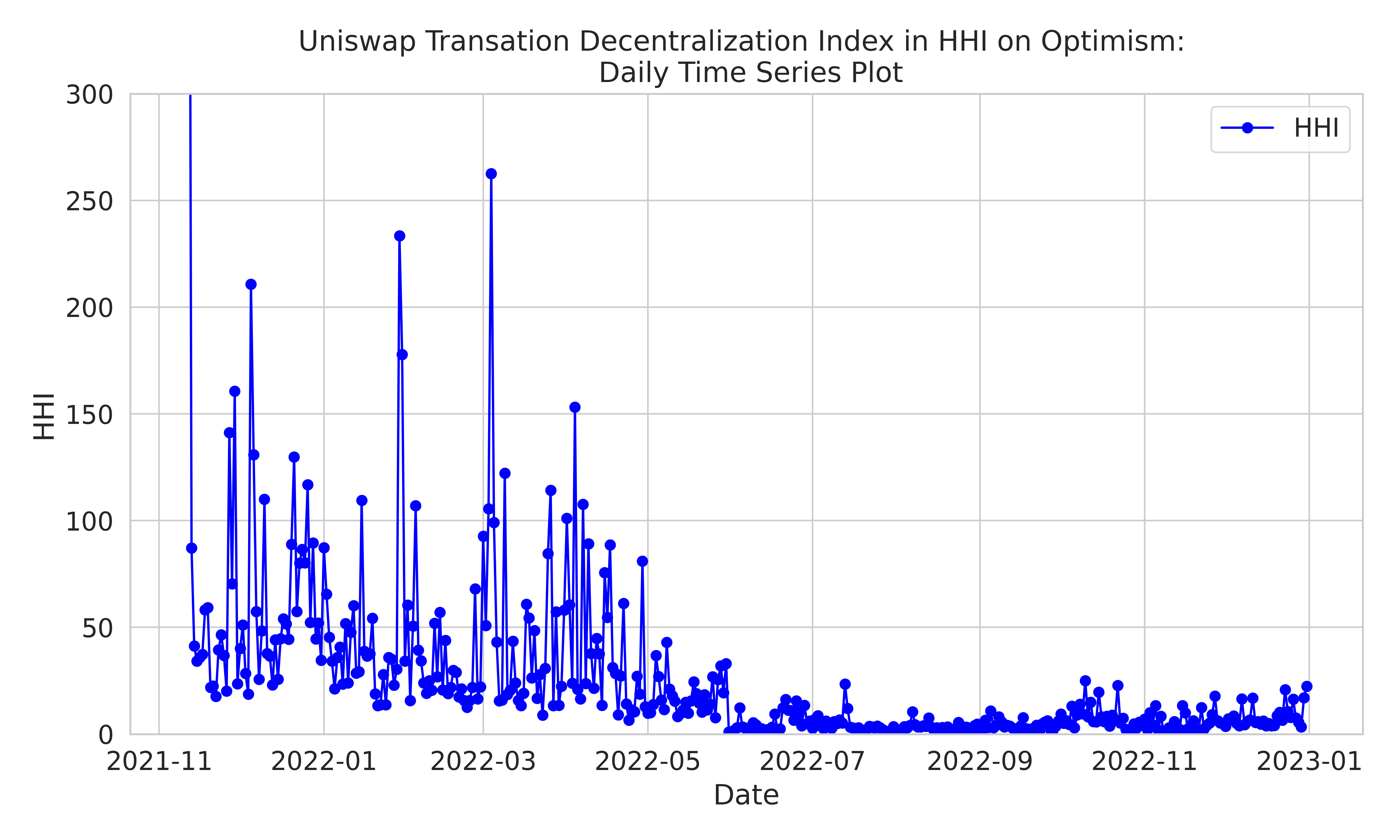}
  \caption{HHI on Optimism }
  \label{estcomp}
\end{subfigure}

\caption{Daily HHI by Network}
\label{fig:HHI}
\end{figure}

\begin{figure}[hbt!]

\begin{subfigure}{.475\linewidth}
  \includegraphics[width=\linewidth]{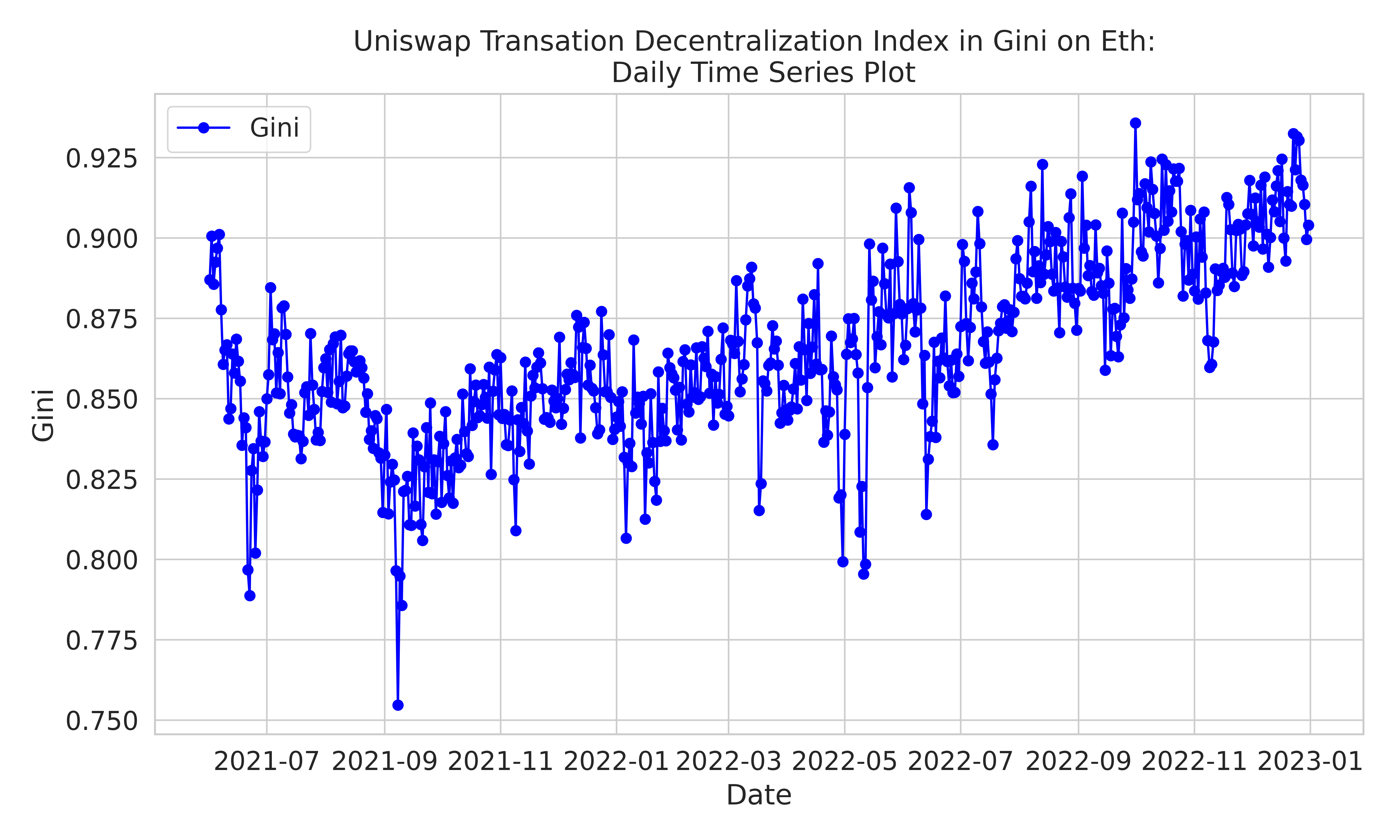}
  \caption{Gini on Ethereum}
  \label{MLEDdet}
\end{subfigure}\hfill 
\begin{subfigure}{.475\linewidth}
  \includegraphics[width=\linewidth]{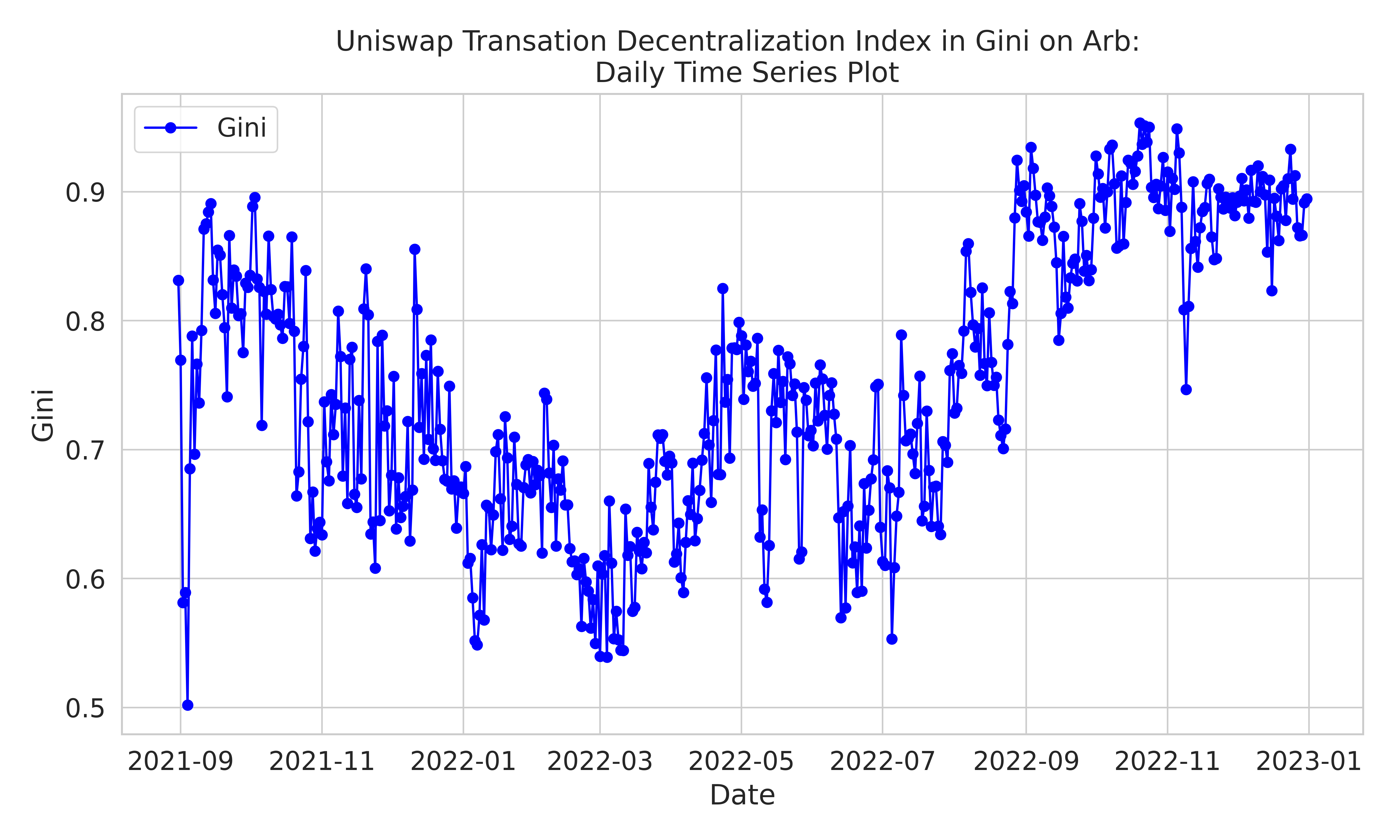}
  \caption{Gini on Arbitrum}
  \label{energydetPSK}
\end{subfigure}

\medskip 
\begin{subfigure}{.475\linewidth}
  \includegraphics[width=\linewidth]{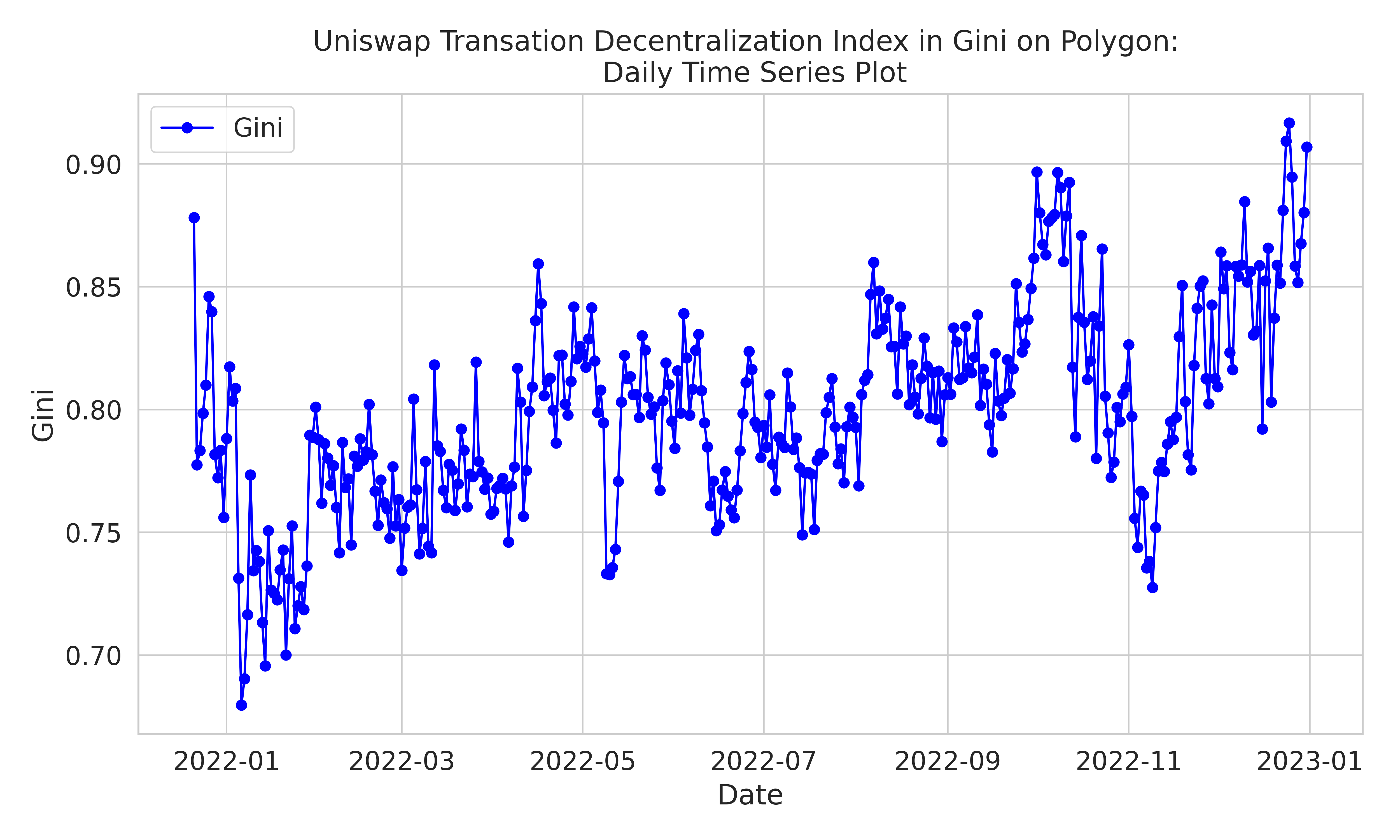}
  \caption{Gini on Polygon}
  \label{velcomp}
\end{subfigure}\hfill 
\begin{subfigure}{.475\linewidth}
  \includegraphics[width=\linewidth]{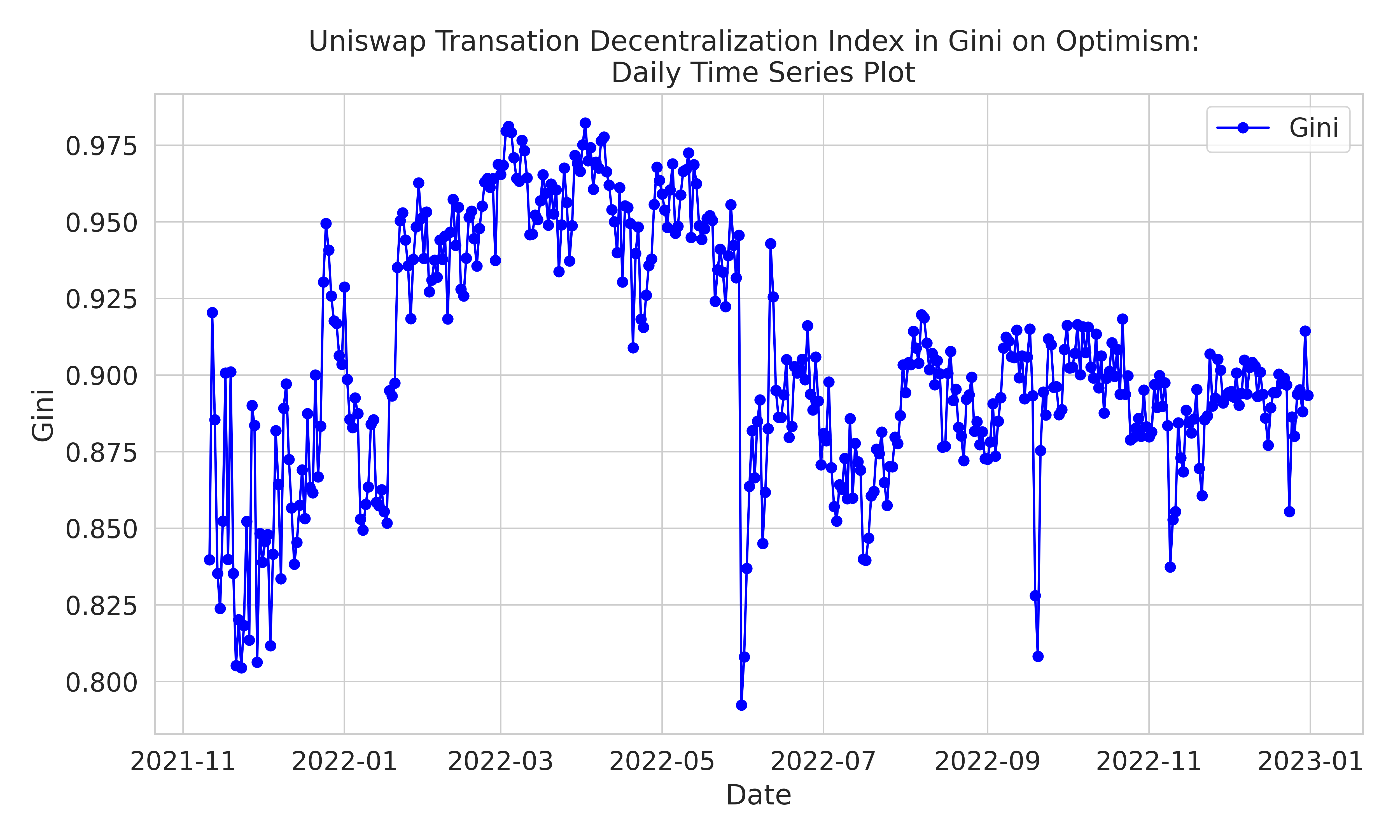}
  \caption{Gini on Optimism }
  \label{estcomp}
\end{subfigure}

\caption{Daily Gini Index by Network}
\label{fig:Gini}
\end{figure}

\begin{figure}[hbtp!]

\begin{subfigure}{.475\linewidth}
  \includegraphics[width=\linewidth]{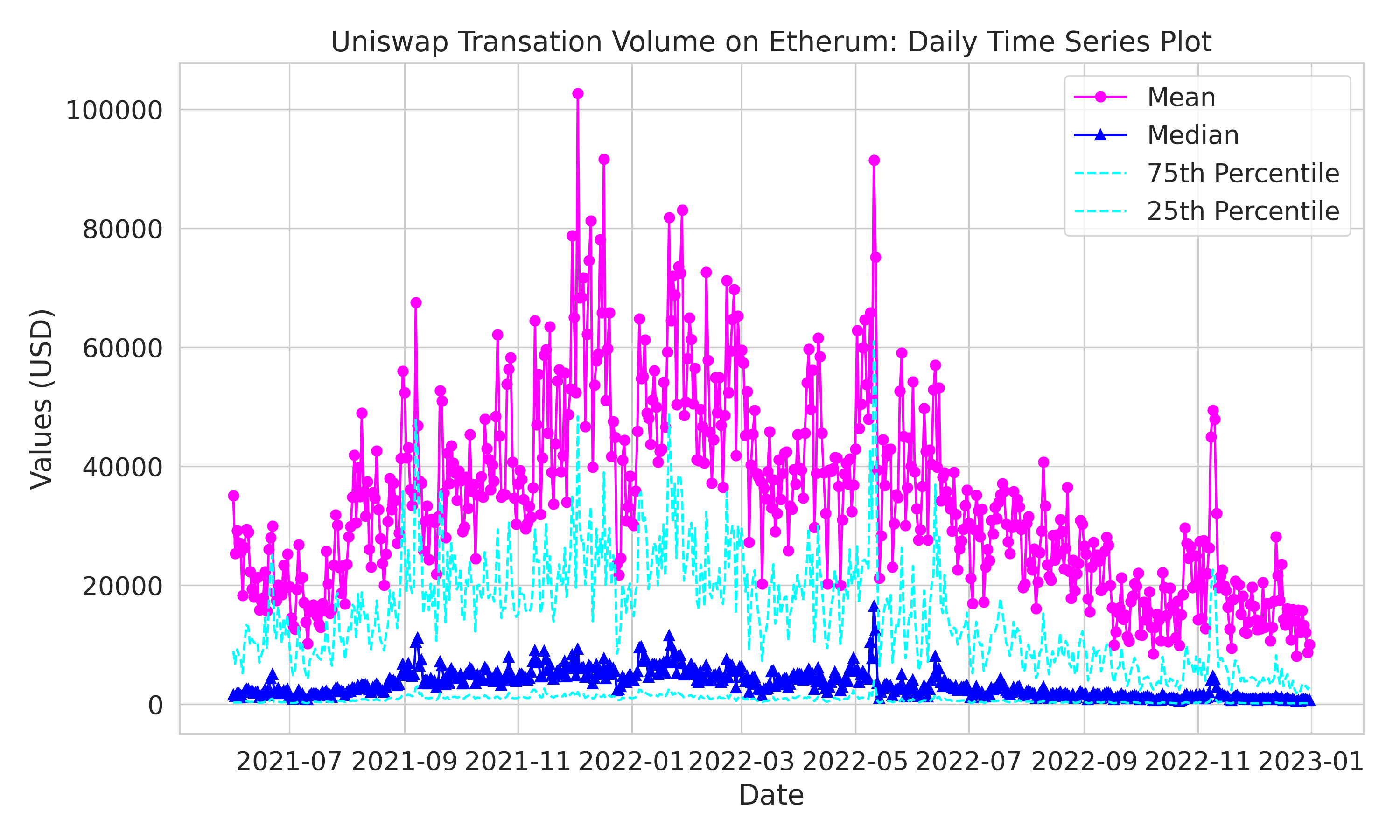}
  \caption{Ethereum Transaction Volume}
  \label{MLEDdet}
\end{subfigure}\hfill 
\begin{subfigure}{.475\linewidth}
  \includegraphics[width=\linewidth]{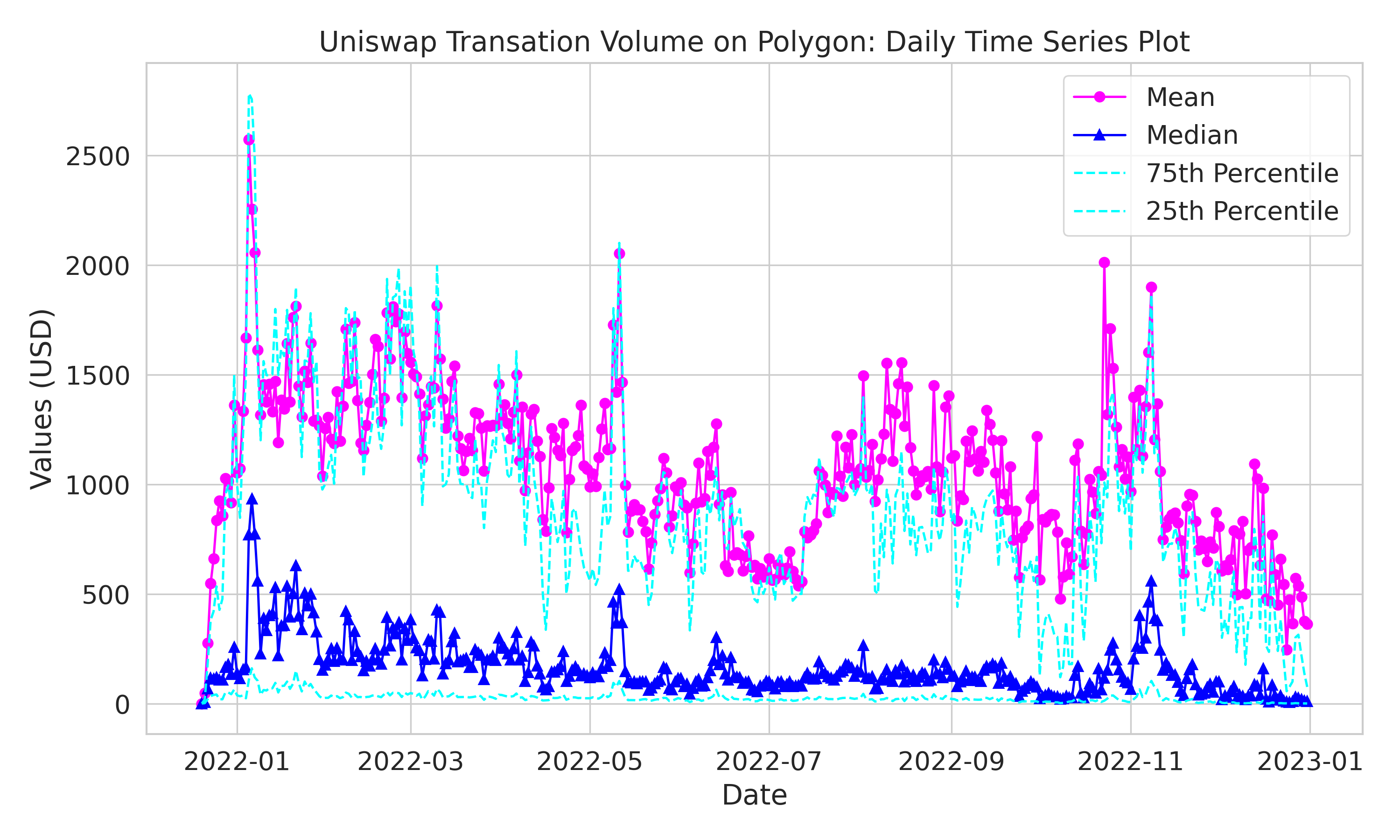}
  \caption{Polygon Transaction Volume}
  \label{energydetPSK}
\end{subfigure}

\medskip 
\begin{subfigure}{.475\linewidth}
  \includegraphics[width=\linewidth]{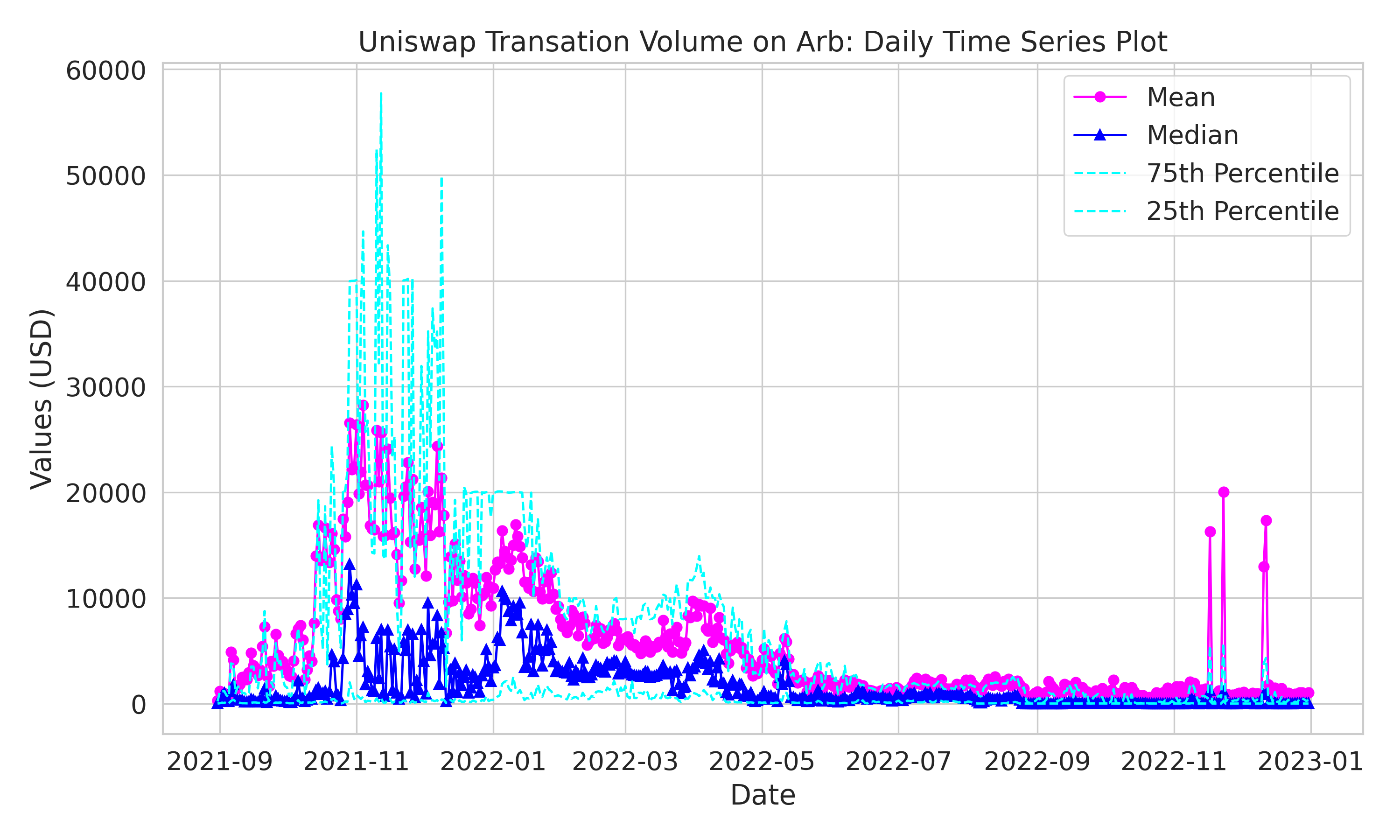}
  \caption{Arbitrum Transaction Volume}
  \label{velcomp}
\end{subfigure}\hfill 
\begin{subfigure}{.475\linewidth}
  \includegraphics[width=\linewidth]{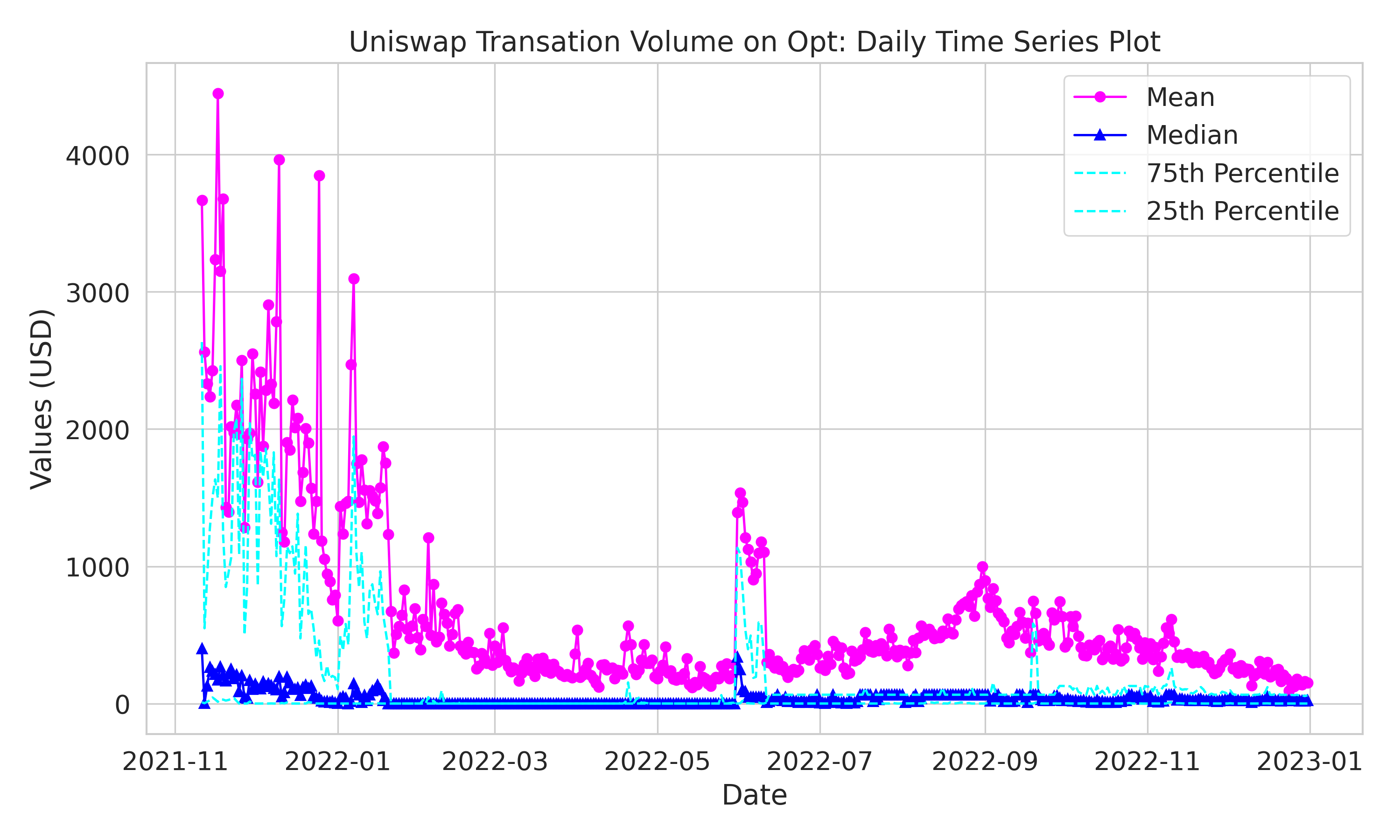}
  \caption{Optimism Transactipn Volume}
  \label{estcomp}
\end{subfigure}

\caption{Daily Transaction Volume by Network}
\label{fig:volume}
\end{figure}




%% file: data_dictionary.tex
\begin{table}[!htbp] \centering
  \caption{Data Dictionary}
\begin{tabular}{@{}rr@{}} 
    \toprule
        \multicolumn{2}{p{9cm}}{\centering Transaction Dataset}  \\
    \midrule
        \multicolumn{1}{p{3cm}}{\raggedright Column Name}  & 
      \multicolumn{1}{p{6cm}}{\raggedright Annotation}\\
    \midrule
      \multicolumn{1}{p{3cm}}{\raggedright ID}  & 
      \multicolumn{1}{p{6cm}}{\raggedright Transaction hash + "\#" + index in swaps Transaction array}\\
      \multicolumn{1}{p{3cm}}{\raggedright Transaction ID}  & 
      \multicolumn{1}{p{6cm}}{\raggedright Pointer to transaction}\\
      \multicolumn{1}{p{3cm}}{\raggedright Timestamp}  & 
      \multicolumn{1}{p{6cm}}{\raggedright Timestamp of transaction}\\
      \multicolumn{1}{p{3cm}}{\raggedright Pool}  & 
      \multicolumn{1}{p{6cm}}{\raggedright Pool swap occured within}\\
      \multicolumn{1}{p{3cm}}{\raggedright Origin}  & 
      \multicolumn{1}{p{6cm}}{\raggedright Txn origin}\\
      \multicolumn{1}{p{3cm}}{\raggedright Token 0}  & 
      \multicolumn{1}{p{6cm}}{\raggedright Token 0 hash}\\
      \multicolumn{1}{p{3cm}}{\raggedright Token 1}  & 
      \multicolumn{1}{p{6cm}}{\raggedright Token 1 hash}\\
      \multicolumn{1}{p{3cm}}{\raggedright Amount 0}  & 
      \multicolumn{1}{p{6cm}}{\raggedright Delta of token0 swapped}\\
      \multicolumn{1}{p{3cm}}{\raggedright Amount 1}  & 
      \multicolumn{1}{p{6cm}}{\raggedright Delta of token1 swapped}\\
      \multicolumn{1}{p{3cm}}{\raggedright Amount USD}  & 
      \multicolumn{1}{p{6cm}}{\raggedright Derived info}\\
    \toprule
        \multicolumn{2}{p{9cm}}{\centering Uniswap Day Dataset}  \\
    \midrule
        \multicolumn{1}{p{3cm}}{\raggedright Column Name}  & 
      \multicolumn{1}{p{6cm}}{\raggedright Annotation}\\
    \midrule
      \multicolumn{1}{p{3cm}}{\raggedright date}  & 
      \multicolumn{1}{p{6cm}}{\raggedright Timestamp rounded to current day by dividing by 86400}\\
      \multicolumn{1}{p{3cm}}{\raggedright volumeETH}  & 
      \multicolumn{1}{p{6cm}}{\raggedright Total daily volume in Uniswap derived in terms of ETH}\\
      \multicolumn{1}{p{3cm}}{\raggedright volumeUSD}  & 
      \multicolumn{1}{p{6cm}}{\raggedright Total daily volume in Uniswap derived in terms of USD}\\
      \multicolumn{1}{p{3cm}}{\raggedright txCount}  & 
      \multicolumn{1}{p{6cm}}{\raggedright Number of daily transactions}\\
      \multicolumn{1}{p{3cm}}{\raggedright tvlUSD}  & 
      \multicolumn{1}{p{6cm}}{\raggedright Tvl in terms of USD}\\
      \multicolumn{1}{p{3cm}}{\raggedright feesUSD}  & 
      \multicolumn{1}{p{6cm}}{\raggedright Fees in USD}\\
    \toprule
        \multicolumn{2}{p{9cm}}{\centering Liquidity Pool Dataset}  \\
    \midrule
        \multicolumn{1}{p{3cm}}{\raggedright Column Name}  & 
      \multicolumn{1}{p{6cm}}{\raggedright Annotation}\\
    \midrule
      \multicolumn{1}{p{3cm}}{\raggedright id}  & 
      \multicolumn{1}{p{6cm}}{\raggedright Pool address}\\
      \multicolumn{1}{p{3cm}}{\raggedright createdAtTimestamp}  & 
      \multicolumn{1}{p{6cm}}{\raggedright Creation}\\
      \multicolumn{1}{p{3cm}}{\raggedright createdAtBlockNumber}  & 
      \multicolumn{1}{p{6cm}}{\raggedright Block pool was created at}\\
      \multicolumn{1}{p{3cm}}{\raggedright token0}  & 
      \multicolumn{1}{p{6cm}}{\raggedright Token0 address}\\
      \multicolumn{1}{p{3cm}}{\raggedright token1}  & 
      \multicolumn{1}{p{6cm}}{\raggedright Token1 address}\\
      \multicolumn{1}{p{3cm}}{\raggedright feeTier}  & 
      \multicolumn{1}{p{6cm}}{\raggedright Fee amount}\\
      \multicolumn{1}{p{3cm}}{\raggedright totalValueLockedUSD}  & 
      \multicolumn{1}{p{6cm}}{\raggedright tvl USD}\\
     \bottomrule
    \end{tabular}
\label{tab: data_dictionary}
\end{table}